\documentclass[]{aastex63}

\usepackage{booktabs}
\usepackage{amsmath}

\begin{document}
	\title{Multi-wavelength Variation Phenomena of PKS 0735+178 on Diverse Timescale}
	
	\author{Yue Fang}
	
	\author{Qihang Chen}
	
	\author{Yan Zhang}
	
	\author{Jianghua Wu$^{\dag}$}
	\affiliation{Department of Astronomy, Beijing Normal University, 100875, Beijing, China}
	\correspondingauthor{Jianghua Wu}
	\email{jhwu@bnu.edu.cn}
	
	\begin{abstract}
		BL Lac object PKS 0735+178 showed some complex multi-wavelength variation phenomena in the previous studies, especially for its color behavior. Bluer-when-brighter, redder-when-brighter and achromatic behavior were all found to be possible long-term trends of PKS 0735+178. In this work, we collected the long-term multi-wavelength data of PKS 0735+178, and also performed a multi-color optical monitoring on intraday timescale. The intraday variability was detected on one night. On the long timescale, a possible 22-day time lag was found between the $R$ and $\gamma$-ray bands. The results of the cross-correlation analysis exhibited strong correlations between various optical bands on both intraday and long timescale. However, only a mild correlation was found between the long-term $\gamma$-ray and $R$-band light curves, which could be interpreted by the different emission mechanisms of the $\gamma$-ray and optical emissions. PKS 0735+178 showed a significant harder-when-brighter in the $\gamma$-ray band, which is consistent with the observed optical bluer-when-brighter trend on both long-term and intraday timescales. We found that the HWB and BWB trends will be enhanced during the active states, especially for the historical low state. Such phenomenon indicates a special activity-dependent color behavior of PKS 0735+178, and it could be well interpreted by the jet emission model.
	\end{abstract}
	
	
	\section{INTRODUCTION}
	Blazars are the most violently variable class of active galactic nuclei (AGN) \citep{Urry1995}, which have their relativistic jets directed towards the earth. They display violent variability of flux and polarization of non-thermal radiation across the entire electromagnetic spectrum. According to whether there are some broad emission lines, blazars can be further divided into flat-spectrum radio quasars (FSRQs) and BL Lac objects. The broadband spectral energy distributions (SEDs) of blazars exhibit two characteristic bumps. One locates at low frequencies (from radio to the UV or X-ray bands) and is dominated by synchrotron radiation. The other locates at high frequencies (the X-ray and $\gamma$-ray bands) and is interpreted as the inverse Compton scattering processes \citep[e.g.,][]{Ulrich1997,Bottcher2007,Sikora2009}. The color behavior can serve as a tool to investigate the nature of emission of blazars and to probe the physical processes in them. The analysis of color-magnitude trends in the monitoring data of blazars has revealed three common patterns, i.e., bluer-when-brighter (BWB), redder-when-brighter (RWB), and achromatic. Through some comprehensive statistical analysis, BL Lac objects frequently show BWB trend, while FSRQs tend to show RWB trend \citep[e.g.,][]{Gu2006,Wu2011,Gaur2012a,Isler2017}. 
	
	PKS 0735+178 (also known as OI 158, S3 0735+17, 4FGL J0738.1+1742) was identified as a BL Lac object \citep{Carswell1974}. Its J2000 coordinates are RA= 07h38m07.3937s, DEC = +17d42m18.998s, and the redshift is $z = 0.45$ \citep{Rector2001}. In 1999, \cite{Hartman1999} reported $\gamma$-ray detection of this blazar using Energetic Gamma Ray Experiment Telescope, and it was found to be quite steady in X-ray and $\gamma$-ray bands with the continuous monitoring \citep{Bregman1984,Nolan2003}. On the contrary, in the optical and radio bands, PKS 0735+178 showed violent long-term variability in the past decades \citep{Webb1988, Ciprini2007}. Also, some intraday variabilities were reported on its optical history \citep{Sagar2004,Goyal2009}. Through a century long optical light curve, \cite{Fan1997} reported a possible period of $\sim14$ years. \cite{Ciprini2007} found 3 characteristic timescales of about 4.5, 8.5, and 11–13 years.
	
	A bronze neutrino event, IC211208A, with a $>30$\% probability of being of astrophysical origin, was localised to the vicinity of PKS 0735+178 (2.1 deg separation, GCN 31191). Follow-up observations showed that it flared in radio (ATel $\#$15105), optical (ATel $\#$15098), X-rays (ATel $\#$15102) and $\gamma$-rays (ATel $\#$15099) nearly simultaneously, with a probability of $>30\%$ being the astrophysical origin of this neutrino event. During the IC211208A event, PKS 0735+178 reached its brightest state in the $R$ band ($\sim14.2$ mag\footnote{\url{http://herculesii.astro.berkeley.edu/kait/agn/lightcurve_all/lightcurve_J0738+1742_all_psf_natural_group_kait.png}}, ATel $\#$15021), and showed a peak $\gamma$-ray daily flux value of $5^{+2}_{-2}\times10^{-7}\ ph\ cm^{-2}\ s^{-1}$, which was 10 times greater than its average 4FGL-DR2 flux (GCN 31194). \cite{Sahakyan2022} presented a multi-wavelength analysis of PKS 0735+178, for which a radio context was also provided by high angular resolution imaging \citep{Nanci2022,Weaver2022}.
	
	Specially, it seems that PKS 0735+178 didn't follow a simple BWB or RWB trend, and researchers have been arguing about its color behavior for more than two decades. As early as in 2000, \cite{Fan2000} collected 10-year historical optical data of PKS 0735+178 and found that it showed a BWB trend, and the studies of \cite{Rani2010} and \cite{Meng2018} also showed this trend. However, \cite{Sandrinelli2014} found a positive correlation between the $R-H$ color vs. the $H$-band magnitude, with a general trend indicating bluer color for decreasing flux, i.e., a RWB trend. In the $\gamma$-ray band, a comprehensive study of blazars (including PKS 0735+178) illustrated a significant harder-when-brighter (HWB) trend for FSRQs, while the BL Lacs showed no propensity toward a HWB or softer-when-brighter trend (SWB) \citep{Williamson2014}. Achromatic behavior or weak correlation in optical bands also appeared in the previous studies on PKS 0735+178 \citep{Gu2006,Ciprini2007}. Recently, \cite{2021PASP..133g4101Y} found that PKS 0735+178 showed two opposite spectral behaviors among different bands, i.e., BWB for spectral indices (SIs) vs. $g$-band flux but RWB for SIs vs. $r$- and $i$-band flux. All these results suggested a complex pattern of its color behavior.
	
	In order to explore the specific color/spectral behaviors of PKS 0735+178, we collected its long-term optical and $\gamma$-ray data, and we also carried out a multi-color optical monitoring program with the 85cm telescope at Xinglong observatory. The descriptions of observations and data reductions could be found in Section \ref{Sec2}. In Section \ref{Sec3}, we studied the intraday variability (IDV) of this object and searched for the possible inter-band time lags through the interpolated cross correlation function (ICCF). In Section \ref{Sec4}, we analyzed its color/spectral behavior in detail, including the behaviors under different timescales and different activity states, to figure out the primary variation mechanism. Finally, a summary is given in Section \ref{Sec5}.
	
	\section{OBSERVATION AND DATA REDUCTION}\label{Sec2}
	\subsection{Optical: Intraday}\label{2.1}
	The intraday monitoring was performed with the 85cm telescope at Xinglong Station of the National Astronomical Observatories Chinese Academy of Sciences (NAOC). This telescope uses the prime focus optical design with a focal ratio $f/3.3$. The CCD was a 2048 × 2048 chip with a field of view of $\sim 32.8 \times 32.8$ $arcmin^2$. We observed PKS 0735+178 for 3 nights in Dec. 2020, and $\sim$400 data points were collected. We have listed the relevant observation information in Table \ref{Tab1}.
	
	\begin{deluxetable}{ccccrrrrrrrrrcc}[ht!]
	\tablecaption{Intraday Results of PKS 0735+178 \label{Tab1}}
	\tablewidth{\columnwidth}
	\tablecolumns{15}
	\tablehead{
		\colhead{Julian Date} & \colhead{Date} & \colhead{Filter} & \colhead{No. Of Expo.} & \multicolumn{4}{c}{Enhanced $F$-test} & \multicolumn{5}{c}{ANOVA test} & \colhead{Var?} & \colhead{Amp} \\
		\cmidrule[0.03cm](r){5-8}\cmidrule[0.03cm](r){9-13} (MJD)& (ISO) & & & $\nu_1$ & $\nu_2$ & $F$ & $F_c$ & $\nu_1$ & $\nu_2$ & $F$ & $F_{star}$ & $F_c$ &  &(\%)\\
		(1) & (2) & (3) & (4) & (5) & (6) & (7) & (8) & (9) & (10) & (11) & (12) & (13) & (14) & (15)}
	\startdata
59188 &  2020-12-04 &  B &  69 &  68 &  340 &  1.57 &  1.51 &  12 &  56 &  7.87 &  0.33 &  2.52 &  V &  $10.6\pm4.7$ \\
&             &  V &  69 &  68 &  340 &  1.80 &  1.51 &  12 &  56 &  6.09 &  0.38 &  2.52 &  V &  $10.1\pm4.8$ \\
&             &  R &  69 &  68 &  340 &  1.99 &  1.51 &  12 &  56 &  5.66 &  0.74 &  2.52 &  V &  $10.1\pm3.8$ \\
59190 &  2020-12-06 &  B &  29 &  28 &  140 &  1.42 &  1.87 &   4 &  24 &  4.07 &  1.21 &  4.22 &  N &          \\
&             &  V &  29 &  28 &  140 &  0.43 &  1.87 &   4 &  24 &   0.79 &  0.64 &  4.22 &  N &          \\
&             &  R &  29 &  28 &  140 &  0.35 &  1.87 &   4 &  24 &   1.41 &  1.33 &  4.22 &  N &          \\
59191 &  2020-12-07 &  B &  32 &  31 &  155 &  0.11 &  1.81 &   5 &  26 &   0.70 &  0.58 &  3.82 &  N &          \\
&             &  V &  32 &  31 &  155 &  0.16 &  1.81 &   5 &  26 &   1.32 &  0.60 &  3.82 &  N &          \\
&             &  R &  32 &  31 &  155 &  0.23 &  1.81 &   5 &  26 &   0.38 &  0.36 &  3.82 &  N &          \\	   
	\enddata
	\tablecomments{The columns are (1) observational date (MJD), (2) observational date (International Organization for Standardization, ISO), (3) filter, (4) number of exposures, (5) the degree of freedom within group in the enhanced $F$-test (the number of values within each group that are free to vary, and it is calculated by $\nu_1=N_b-1$, where $N_b$ is the number of data points in the blazar's light curve), (6) the degree of freedom between groups in the enhanced $F$-test (the number of values between all the groups that are free to vary, and it is calculated by $\nu_2=(\sum_{j=1}^{k} N_{j})-k$, where $N_j$ is the number of data points in the $j$th comparison star's light curve and $k$ is the number of comparison stars), (7-8) $F$ and the critical value $F_c$ in the enhanced $F$-test, (9-10) two degrees of freedom in the ANOVA test (their meanings are the same as those in the enhanced $F$-test, but calculated by $\nu_1=g-1$ and $\nu_2=N_b-g$, respectively, where $g$ is the number of groups), (11-13) $F$ of PKS 0735+178, $F$ of star C and $F_c$ in the ANOVA test, (14) variable or not, (15) variability amplitude and associated uncertainty, respectively.}
\end{deluxetable}

\begin{figure}[hb!]
		\begin{center}
			\includegraphics[angle=0,scale=0.5]{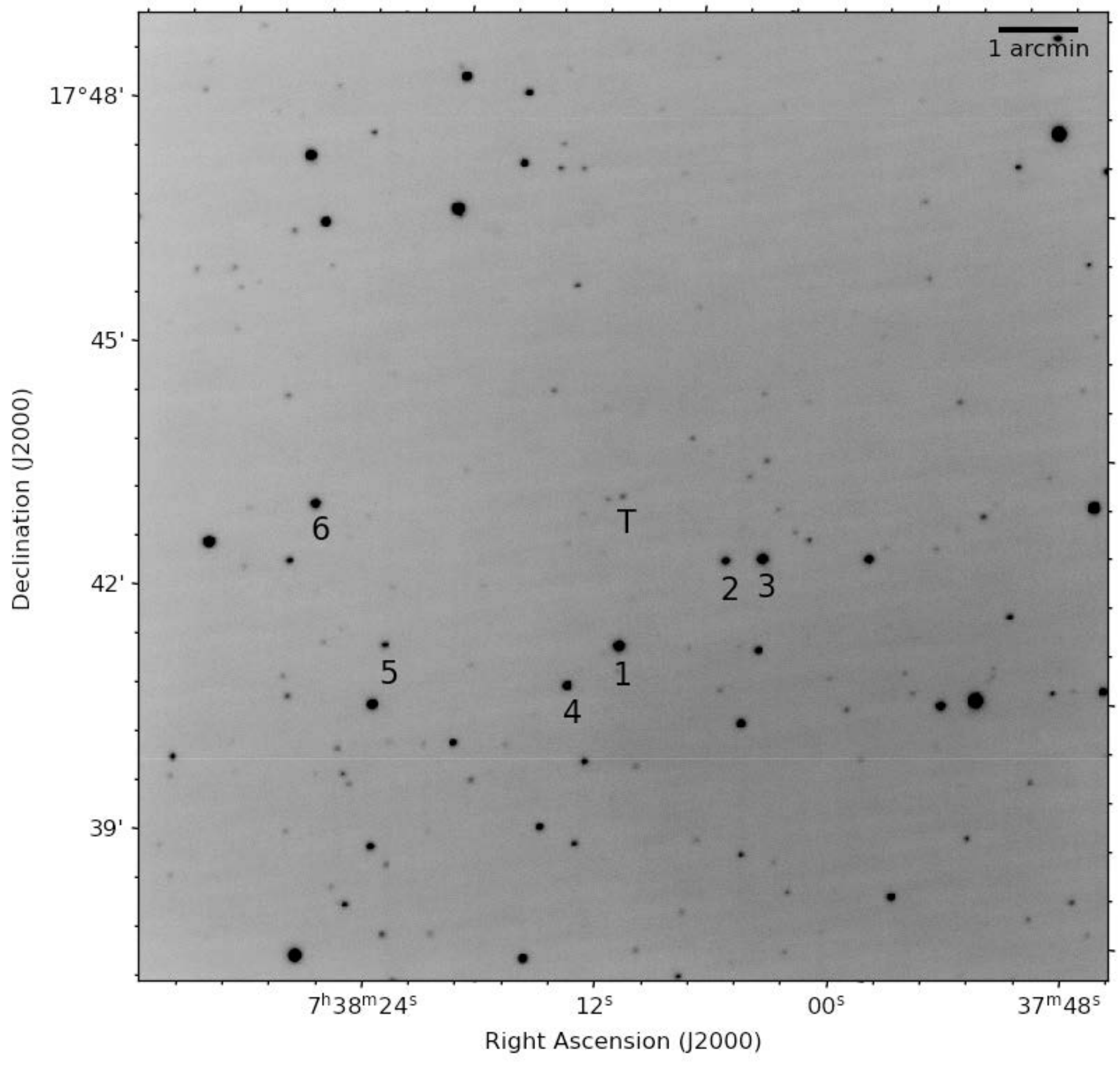}
			\caption{Finding chart of PKS 0735+178 in the $R$ band. The labels ``T" and ``1-6" represent PKS 0735+178, the comparison star, the check star, and four field stars, respectively.\label{finding}}
		\end{center}
	\end{figure}

	As shown in Fig.~\ref{finding}, PKS 0735+178 and six field stars are labeled. The data reduction followed the standard process including bias-subtraction, flat-fielding and instrument magnitudes extraction with the Image Reduction and Analysis Facility (IRAF)\footnote{IRAF is distributed by the National Optical Astronomy Observatories, which are operated by the Association of Universities for Research in Astronomy, Inc., under cooperative agreement with the National Science Foundation.} software. Through aperture photometry, we extracted the instrument magnitudes of PKS 0735+178 and six field stars. In order to obtain the best aperture, the photometry was carried out with ten different aperture radii range from 1 to 3 times of the full width at half-maximum (FWHM), and the inner and outer radii of the sky annulus were accordingly set as 6 and 8 times of the FWHMs, respectively. We finally selected 1.6$\times$FWHM, which gave minimum standard deviations of the differential magnitudes between the check star and the comparison star. In order to minimize the error of calibration, star 1 was chosen as the comparison star, which is the brightest one of all six. Moreover, the closest star 2 was chosen as the check star. The fluxes of PKS 0735+178 and check star were calibrated with respect to a comparison star. The calibrated magnitudes of our target and star 2 were presented in Fig.~\ref{IDLC}. For clarity, the $B$- and $V$-band light curves are shifted by $-0.3$ mag and $-0.4$ mag, respectively. Similar shifts of $-1.1$ mag and $-0.5$ mag were also applied respectively to the $B$- and $R$-band light curves in the small panels.
	
	\subsection{Optical: Long-term}
	We collected the historical optical data from various archives, including those of the Steward Observatory \citep{Smith2009}\footnote{\url{http://james.as.arizona.edu/~psmith/Fermi}}, the Small and Moderate Aperture Research Telescope System (SMARTS)\footnote{\url{www.astro.yale.edu/smarts}}, the Catalina Real-Time Transient Survey (CRTS) \citep{Drake2009}\footnote{\url{http://crts.caltech.edu/}}, the Katzman Automatic Imaging Telescope (KAIT)\footnote{\url{http://herculesii.astro.berkeley.edu/kait/agn/}} and the All-Sky Automated Survey for Supernovae \citep{Jayasinghe2019}\footnote{\url{http://www.astronomy.ohio-state.edu/asassn}}. Notice that the data from SMARTS are actually raw images after bias subtracted and flat field calibrated, which can be queried from NOIRLab Astro Data Archive\footnote{\url{https://astroarchive.noao.edu}}, and the subsequent processing is the same as in Sec.~\ref{2.1}. The collected data are in the $V$-, $R$-, and $I$-bands, but we will not consider the $I$-band due to the low sampling. If more than one data point is present in a night, the average magnitudes are taken.
	
	
	\subsection{$\gamma$-ray}
	The Fermi large area telescope (LAT) data of PKS 0735+178 from 2009 January 1 to 2019 December 31 were downloaded, and a circular region of interest (ROI) of $20^{\circ}$ was chosen. The energy range was restricted to 0.1-200 GeV. We performed the unbinned likelihood analysis using version v2.0.8 of the Fermi Science Tools, the background models from the $\text{\textbf{iso\_P8R3\_SOURCE\_V3\_v1.txt}}$ isotropic template, and the $\text{\textbf{gll\_iem\_v07.fits}}$ Galactic diffuse emission model. We utilized analysis cuts of $\texttt{evclass = 128}$, $\texttt{evtype = 3}$,  and $\texttt{zmax=90}$ of the photon data. We generated the XML model files using $\textbf{make4FGLxml.py}$, and all the spectral indices and fluxes of the  sources were fixed, other than the target, at their 4FGL values. We considered one measurement to be a successful detection when the test statistics (TS) value exceeded 10, which corresponds to a $3\sigma$ criteria. The light curve was generated with a 15-days bin, and we used a power law (PL) function, i.e., $dN(E)/dE=N_0 (E/E_0)^{-\Gamma_\gamma}$, to fit the $\gamma$-ray fluxes in eight logarithmically equal energy bins on each time bin. The variation behavior of spectra will be studied by pairing the simultaneous flux and SI $\Gamma_\gamma$. The long-term light curves in all three bands were presented in Fig.~\ref{LTLC}. Moreover, the measurements with TS $ <10$ were labeled by ``upper limits" (red triangles).

	\section{Variations}\label{Sec3}
	\subsection{Light Curves}\label{Sec3.1}
\begin{figure}[ht!]
		\begin{center}
			\includegraphics[angle=0,scale=0.42]{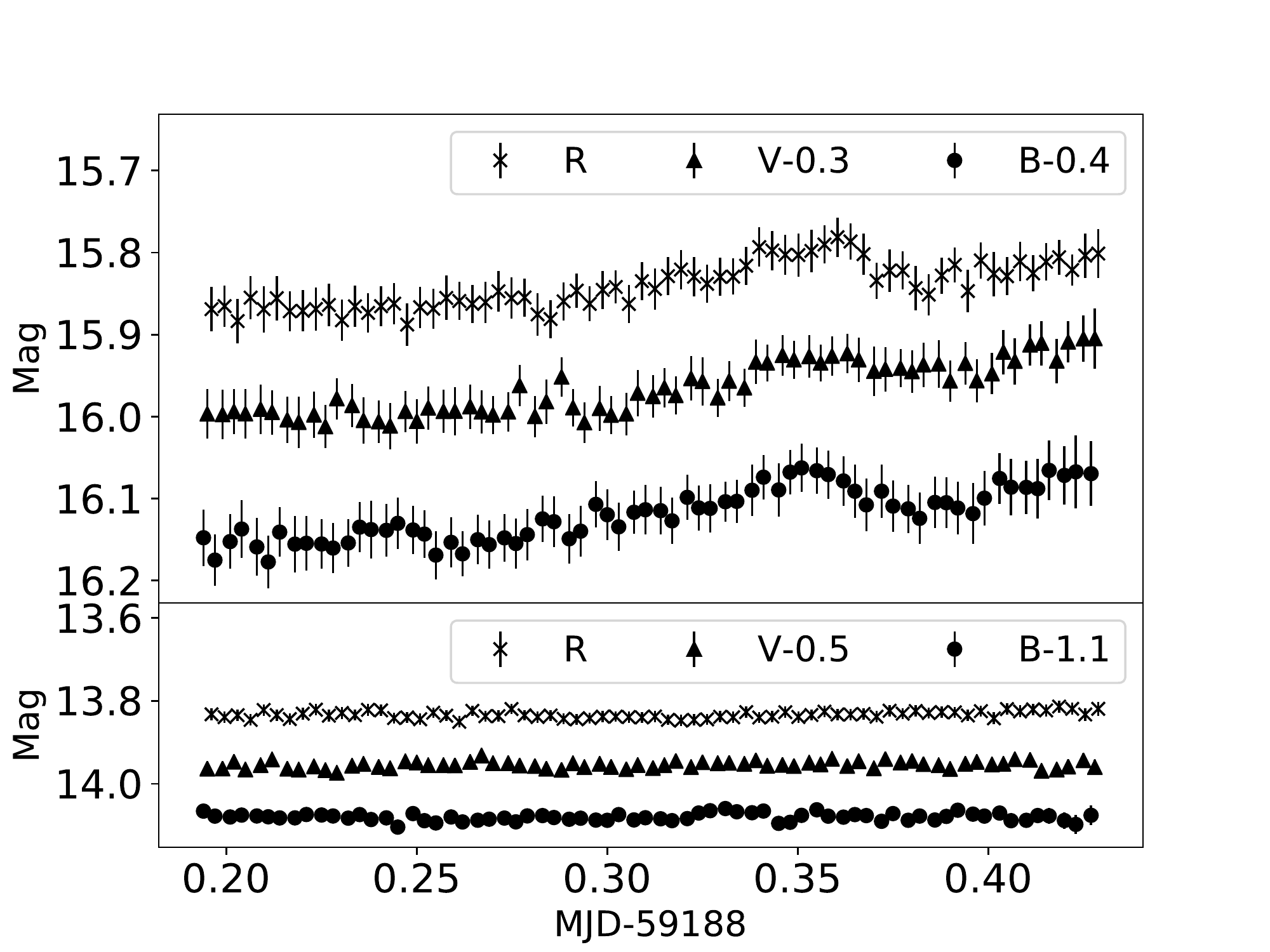}
			\includegraphics[angle=0,scale=0.42]{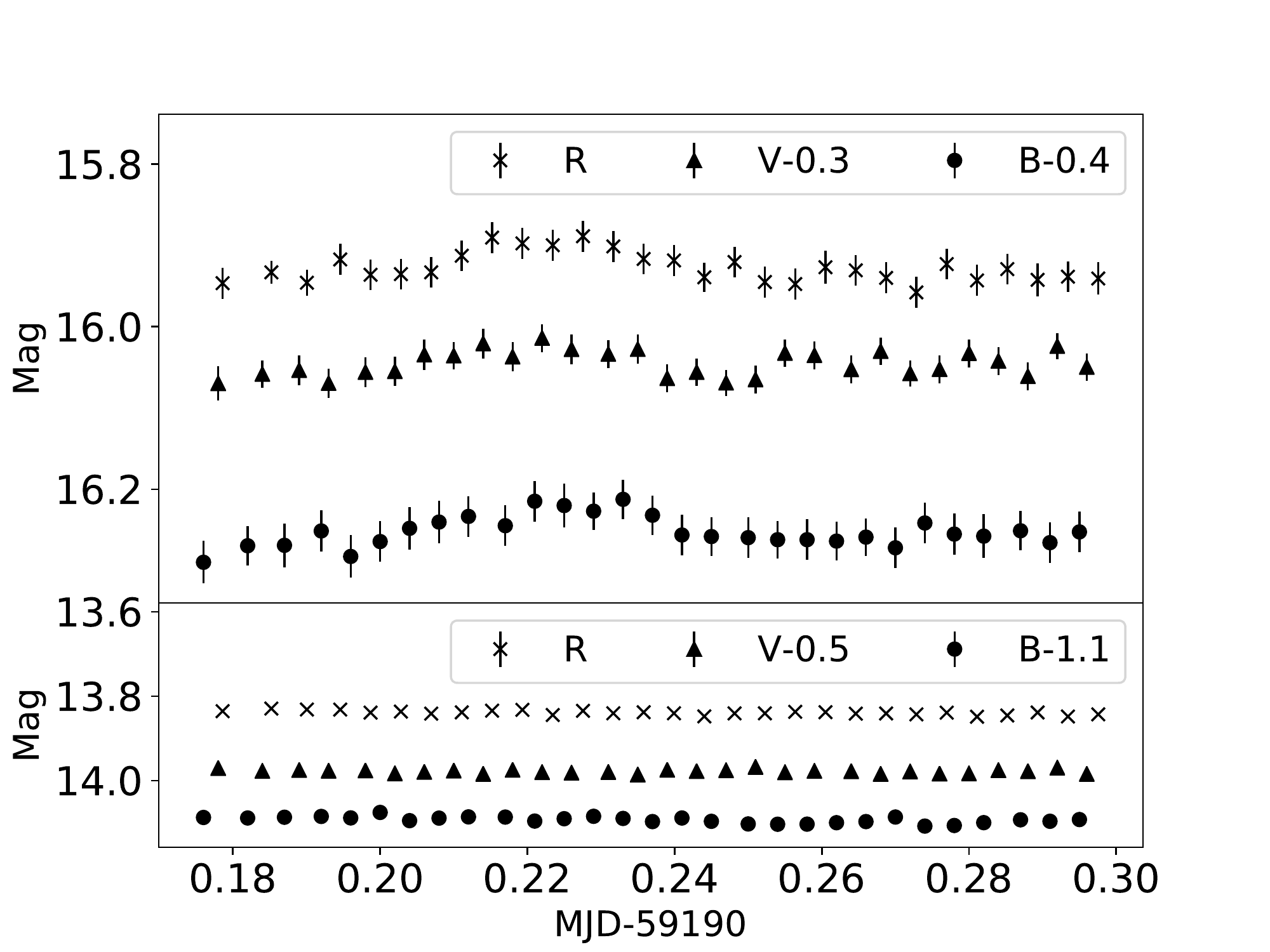}
			\includegraphics[angle=0,scale=0.42]{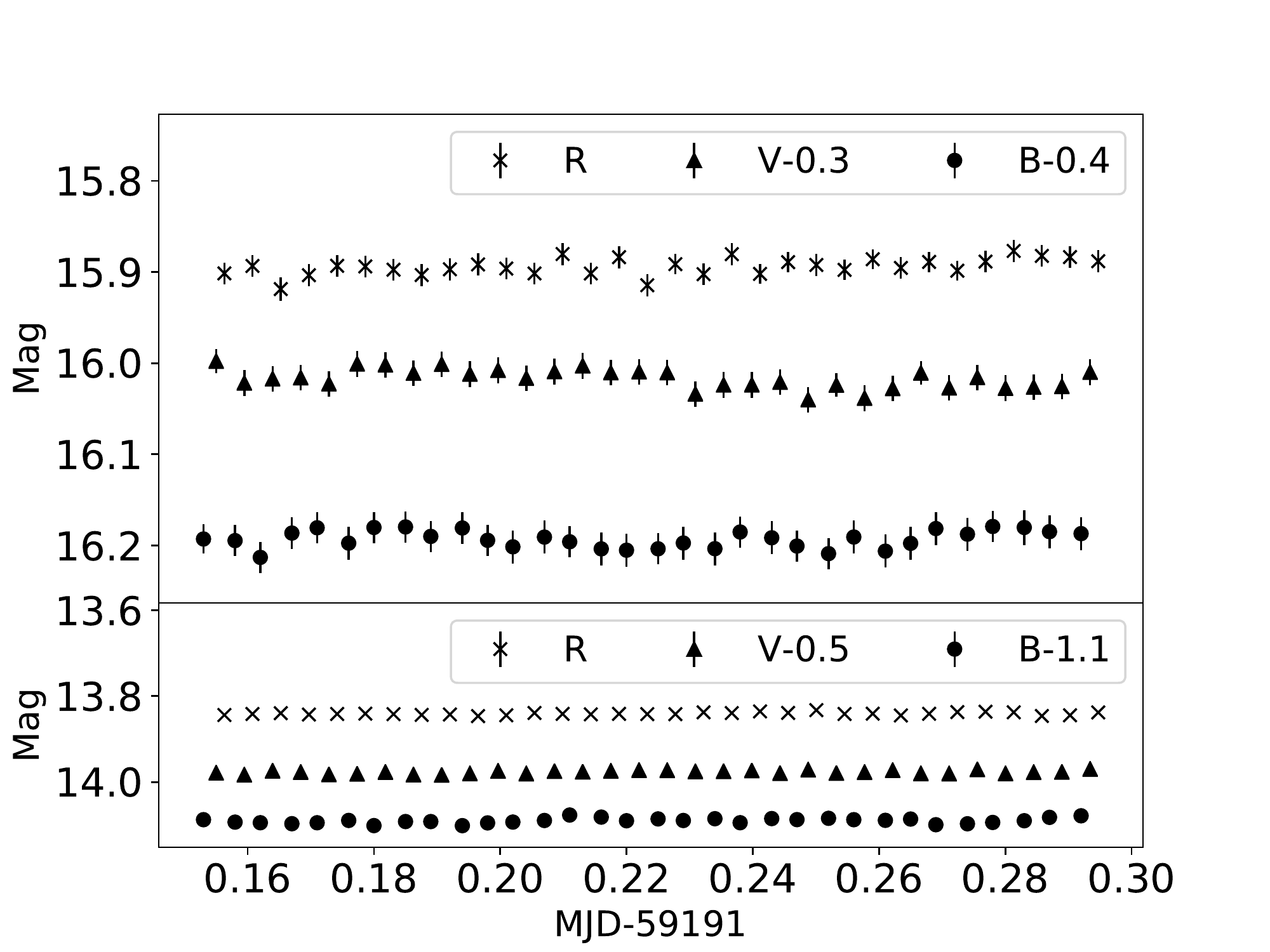}
			\caption{The intraday light curves of PKS 0735+178 and the check star are given in the large and small panels, respectively. For clarity, the $B$- and $V$-band light curves of PKS 0735+178 are shifted by $-0.4$ and $-0.3$ mag, respectively, while those of the check star are shifted by $-1.1$ and $-0.5$ mag, respectively. \label{IDLC}}
		\end{center}
	\end{figure}

\begin{figure}[ht!]
		\begin{center}
			\includegraphics[scale=0.5]{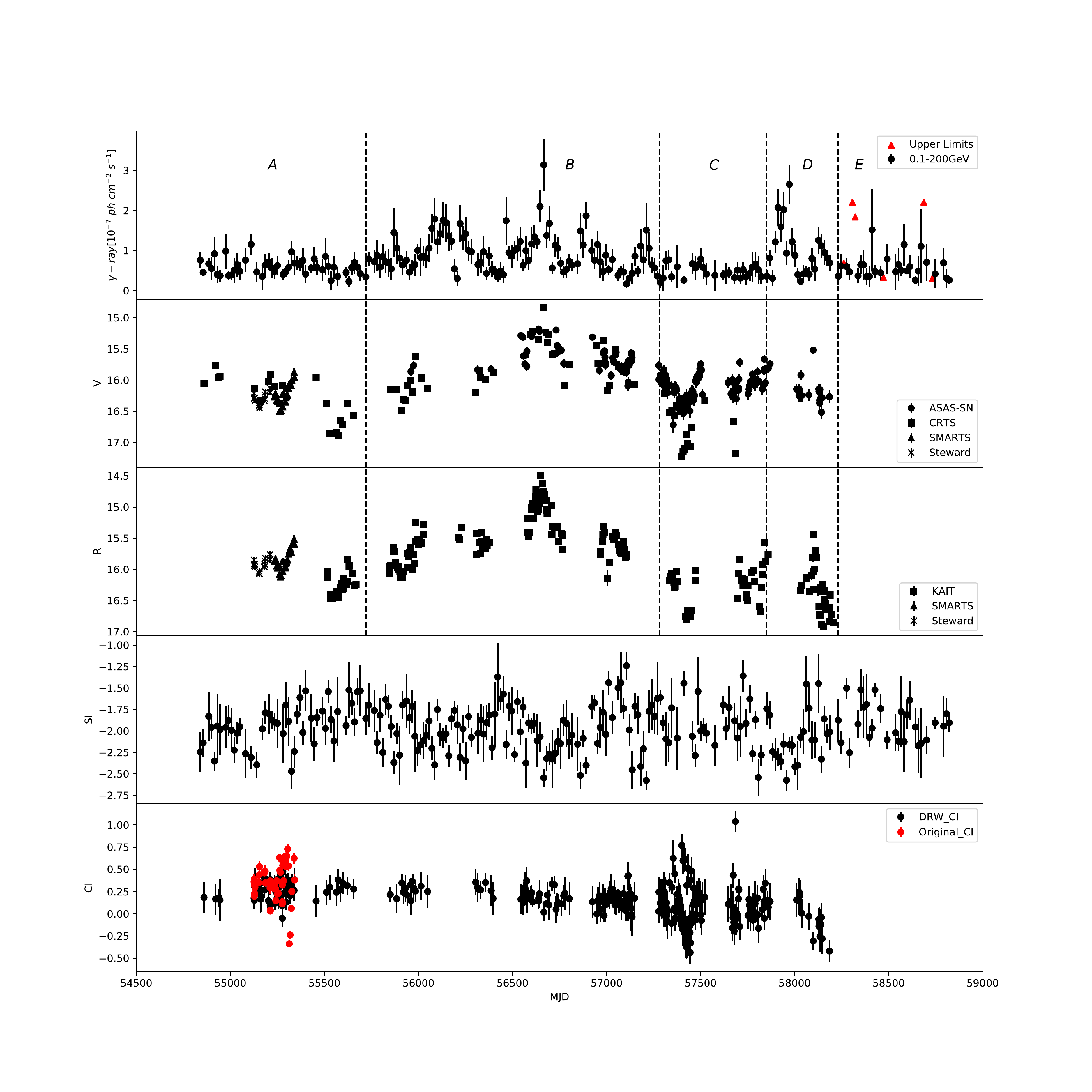}
			\caption{Top panel: the light curve of $\gamma$-ray in the range of 0.1-200 GeV, and the red triangles represent the upper limits of the measurements with TS $<10$. The dashed vertical lines divide the light curve into five phases according to the activities (see Section \ref{Sec3.1} for detail). The second and third panels: the $V$- and $R$-band light curves, which were segmented according to the phases in the $\gamma$-ray band. The fourth panel and bottom panel are the SI of the $\gamma$-ray and $V-R$ CI, respectively. Red dots represent those CIs obtained with the original magnitude pairs without damped random walk interpolation. \label{LTLC}}
		\end{center}
	\end{figure}

	From visual inspection of Figure \ref{IDLC} and \ref{LTLC}, we can find that the optical light curves are well correlated with each other on both intraday and long timescales, and a significant IDV is exhibited on MJD 59188. However, on the long timescale, the correlation between the optical band and the $\gamma$-ray band is clearly weaker than that between two optical bands. A prominent flare was observed during MJD 56500-56800 in the $V$ and $R$ bands. A historical low state (17.23 mags) was observed in the $V$ band during MJD 57350-57450, which was $\sim$0.2 mag fainter than the previous lowest state ($\sim$17 mags during MJD 50400-50600, see \cite{Ciprini2007}). During MJD 59188-59191, the source was the faintest on MJD 59190 with 15.95 mags in the $R$ band and reached the brightest state on MJD 59188 with 15.78 mags. During our long-term optical observations, PKS 0735+178 reached the brightest state on MJD 56649 in the $R$ band, which was $\sim0.3$ magnitude fainter than its historical brightest state on MJD 59551, i.e., six days before the neutrino event IC211208A. We noticed that all the optical flares over the past decades have far lower intensity than the most recent flare possibly associated with the neutrino event IC211208A. It brightened by $\sim$1 mag in 50 days since MJD 59500, and then dimmed at nearly the same rate (GCN 31529), which was about three times the rate of that flare during MJD 56500-56800. In the $\gamma$-ray band, the object is 44 times brighter in the highest state than in the lowest state.

	
	In order to determine the variability on intraday timescale, we used two robust statistical tests, namely, the power-enhanced version of $F$-test \citep{Diego2014} and analysis of variance (ANOVA) \citep{Diego1998}, which were widely used in the past IDV studies \citep[e.g.][]{Gaur2015,Polednikova2016,Meng2017,Zhang2018,Pandey2020}. The detailed algorithms can be found in the corresponding references. If the $F$ exceeds the critical value $F_c$ at the $99\%$ confidence level, the null hypothesis will be rejected and the existence of variability (V) will be confirmed, otherwise it is non-variable (N). To avoid detecting spurious variability, an additional ANOVA test was performed to the check star. Only both tests of PKS 0735+178 show ``V" and the test of check star is ``N", PKS 0735+178 is genuinely variable (``V"), otherwise it is non-variable (``N"). The final results were listed in Table.~\ref{Tab1}, including two corresponding degrees of freedom, the $F$ value and the critical value $F_c$. PKS 0735+178 only showed IDV on MJD 59188. We found that the observing duration on MJD 59188 was the longest among three nights, and most of the variabilities on this night occurred in the second half night. Longer sampling would naturally have higher probability to detect IDV.
	
	We also calculated the variation amplitudes on this night, by
	\begin{equation}\label{Amp}
	Amp = 100\%\times \sqrt{(A_{max}-A_{min})^2 - 2 \sigma^2},
	\end{equation}
	where $A_{max}$ and $A_{min}$ are the maximum and minimum magnitudes, respectively, and $\sigma$ is the measurement error \citep{Heidt1996}. The propagating uncertainty of $Amp$ is given by $\sigma_{Amp}=(\frac{A_{max}-A_{min}}{Amp})\cdot\sqrt{(\sigma_{A_{max}}^2+\sigma_{A_{min}}^2)}$, where the $\sigma_{A_{max}}$ and $\sigma_{A_{min}}$ are the measurement errors corresponding to $A_{max}$ and $A_{min}$, respectively. The amplitudes and associated uncertainties were also given in Table \ref{Tab1}. We can find that the variation amplitude tends to decrease with decreasing frequency, which is consistent with the common tendency, i.e., the higher the energy band, the greater the IDV amplitude \citep{Webb1998,Meng2017}.
	
	On the long-term trend, three distinct $\gamma$-ray flares were observed during MJD 55950-56420, 56420-56780 and 57840-58040, respectively, and the second flare was observed quasi-simultaneously in the $V$ and $R$ bands. According to the activities, we divided the $\gamma$-ray light curve into five phases, namely, phase $A$ (the 1st quiescent, and from MJD 54839 to 55720), phase $B$ (the 1st active state, and from MJD 55720 to 57280), phase $C$ (the 2nd quiescent, and from MJD 57280 to 57850), phase $D$ (the 2nd active state, and from MJD 57850 to 58230) and phase $E$ (the 3rd quiescent, and from MJD 58230 to the end), which were distinguished by vertical dashed lines in Fig.~\ref{LTLC}. The optical light curves were segmented according to the $\gamma$-ray phase.

	\begin{figure}[ht!]
		\begin{center}
			\includegraphics[scale=0.5]{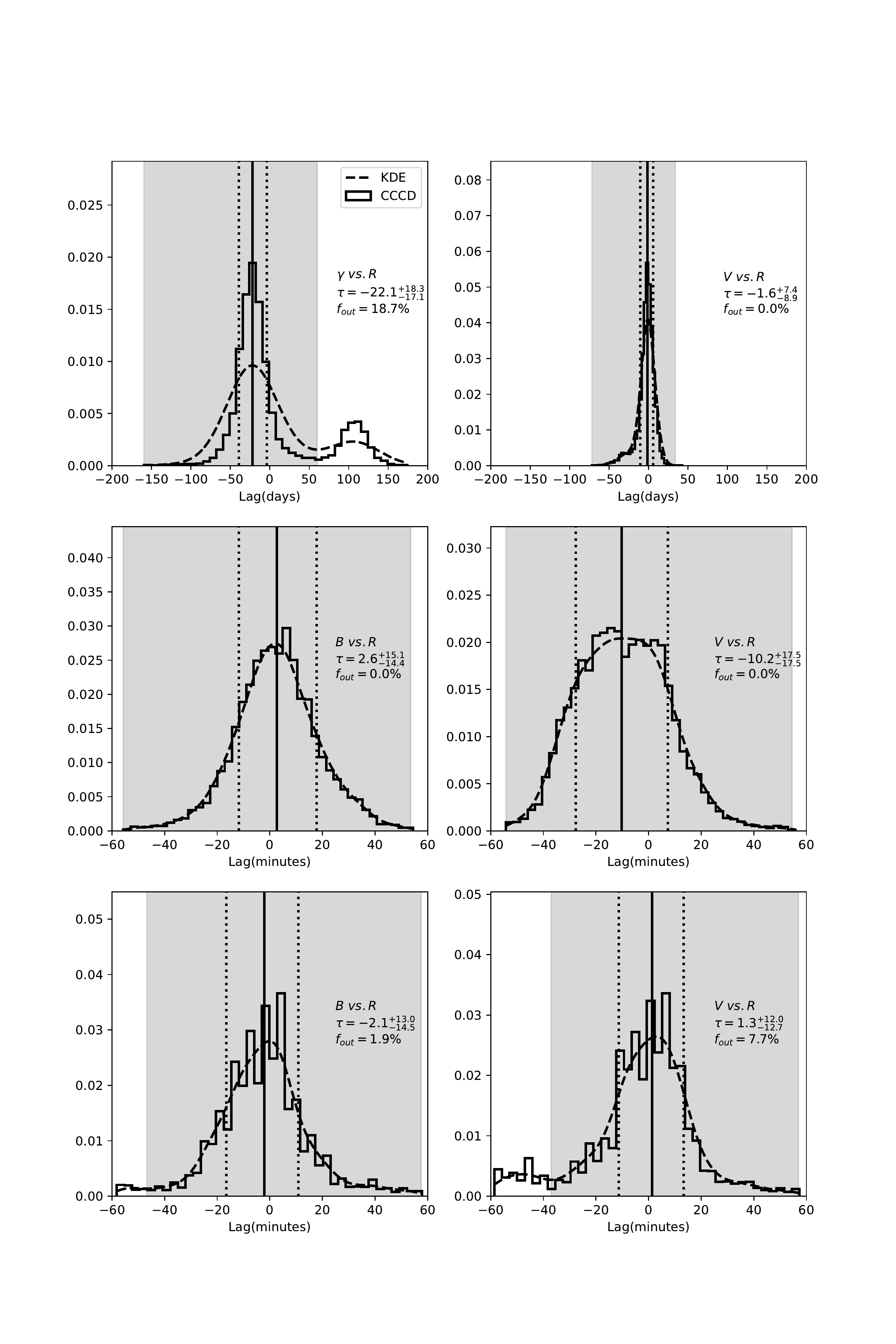}
			\caption{The rows from top to bottom give the lag measurements on MJD 54839-58821, MJD 59188 and MJD 59190, respectively. The measurements on MJD 59191 were excluded due to the too small $r_{max}$ (see Table \ref{Tab2}). The solid and dotted lines indicate the measured lags and uncertainties, respectively. The corresponding lag calculation results are listed in each panel. The dashed lines indicate the smoothing of the weighted CCCDs and the histograms are the unweighted CCCDs. The gray regions highlight the ranges of lags included in the final calculation, and the removal efficiency $f_{out}$ were also given. \label{lag}}
		\end{center}
	\end{figure}
	
	\subsection{Cross-Correlation Analysis}
	We performed the correlation analysis to search for the possible inter-band time lags by using ICCF\footnote{The public available algorithm \texttt{PyCCF} can be found in \url{http://ascl.net/code/v/1868}} \citep{Peterson1998, Peterson1998a}. Through shift and linear interpolation, the Pearson coefficient $r$ between the two light curves will be calculated at each given time shift $\tau$. The optimal time lag and its uncertainties were obtained via the flux randomization/random subset sampling (FR/RSS) method, using Monte Carlo (MC) realizations, as discussed in \cite{Peterson2004}. Those realisation with a peak correlation coefficient $r_{peak} < 0.5$ will be labeled as a failed measurement, and the failure rate $f_{fail}$ will be recorded. The centroid lag will be estimated using points surrounding the $r_{peak}$, and we can obtain the cross correlation centroid distribution (CCCD) after 10000 MC realizations. The search range of time lag was empirically fixed as $[-200, 200]$ days and $[-60,60]$ minutes for long-term and intraday timescales, respectively. 
	
	However, the alias of multiple significant peaks has been reported in the previous studies \citep{Grier2017,Zhang2018,Homayouni2019,Li2019}. Thus we followed \cite{Grier2017} and introduced the weighted lag identification. The weight of each data point in CCCD were defined as $P = [N(\tau)/N(0)]^2$, where $N(\tau)$ corresponds to the number of overlapping epochs at a time shift $\tau$. After a Gaussian kernel estimation (KDE), we can obtain the peak with the largest area in CCCD, and the data points beyond the range of this primary peak will be removed. The final lag was estimated as the median of the unweighted \& unsmoothed CCCD within the range of the primary peak, and the 16$^{\rm th}$ and 84$^{\rm th}$ percentiles were chosen as the lower and upper uncertainties, respectively. Empirically, a reliable result should satisfy $f_{fail} \leq 30\%$, $f_{out}\leq 50\%$ (the ratio of the removed data points to the total data points) and $r_{max} \geq 0.5$. All the measurement results that satisfied these criteria were presented in Fig.~\ref{lag}, and corresponding $f_{fail}$, $f_{out}$, the maximum correlation coefficient $r_{max}$ and time lags were listed in Table~\ref{Tab2}. The negative lags indicate that the former band leads the latter one.
	
	On the intraday timescale, all the time lags were near-zero within the error range. A similar result was obtained between the $V$ and $R$ bands on the long timescale. The non-detection of the lag between variations in different optical bands may be because the emitting regions of these optical bands are too close to each other to result in the time lag \citep{Carini2011,Wu2012,Meng2017}. Moreover, according to the data compiled from \cite{Sandrinelli2014}, a near-zero lag ($-3.6^{+4.2}_{-4.2}$ days, see Appendix~\ref{infrared}) was also found between the $R$ and $J$ bands. The only one non-zero result was the 22-day lag between the $R$ and $\gamma$-ray bands. On the long timescale, the $r_{max}$ between the $\gamma$-ray and $R$ bands is only 0.5, which is much less than the $r_{max}=0.82$ between two optical bands. Several studies also found that the correlation between the optical and $\gamma$-ray bands was much weaker than that between various optical bands \citep[e.g.,][]{2013ApJ...763L..11C,2014ApJ...797..137C,2019ApJ...880...32L,2020MNRAS.498.5128R}. This is mainly due to the different origins of the optical and $\gamma$-ray emissions, the former is dominated by the synchrotron radiation while the latter is by the inverse Compton process. Alternatively, \cite{2020MNRAS.498.5128R} demonstrated that the changes of the magnetic field strength could lead to this phenomenon.
	
	\begin{deluxetable}{cccccc}
	\tablecaption{Results of cross-correlation analysis.\label{Tab2}}
	\centering
	\tablewidth{\columnwidth}
	\tablecolumns{6}
	\tablehead{
		\colhead{Date} & \colhead{Bands} & \colhead{$f_{fail}$} & \colhead{$f_{out}$} & \colhead{$r_{max}$} & Lag \\
		& & ($\%$) & ($\%$) & &  }
	\startdata
54839-58821 & $\gamma$-ray vs. $R$ & 0.5  & 18.7 & 0.52 & $-22.1^{+18.3}_{-17.1}$ d \\
			& $V$ vs.  $R$ 		   & 0.2  & 0	 & 0.82 & $-1.6^{+7.4}_{-8.9}$ d \\
	59188   & $V$ vs.  $R$         & 5.6  & 0  & 0.86 &  $-10.2^{+17.5}_{-17.5}$ min \\
		    & $B$ vs.  $R$         & 2.8 & 0 & 0.90 &   $2.6^{+15.1}_{-14.4}$ min \\
	59190   & $V$ vs.  $R$         & 5.5  & 7.7  & 0.62 & $1.3^{+13.0}_{-12.7}$ min \\
		    & $B$ vs.  $R$         & 1.8  & 1.9  & 0.84 & $-2.1^{+13.0}_{-14.5}$ min \\
	59191   & $V$ vs.  $R$         & 27.5  & 44.9 & 0.35 & / \\
		    & $B$ vs.  $R$         & 17.5 & 50.2 & 0.18 & / \\
	\enddata
	\tablecomments{The columns stand for observational date in the unit of MJD, bands, failure rate, removal efficiency, the maximum correlation coefficient and the measured lag through \texttt{PyCCF}. The negative lags indicate that the former band leads the latter one.}
	\end{deluxetable}

	\section{Color/Spectral Behavior}\label{Sec4}
	\subsection{Observational Findings}\label{Sec4.1}
\begin{figure}[ht!]
		\begin{center}
			\includegraphics[scale=0.4]{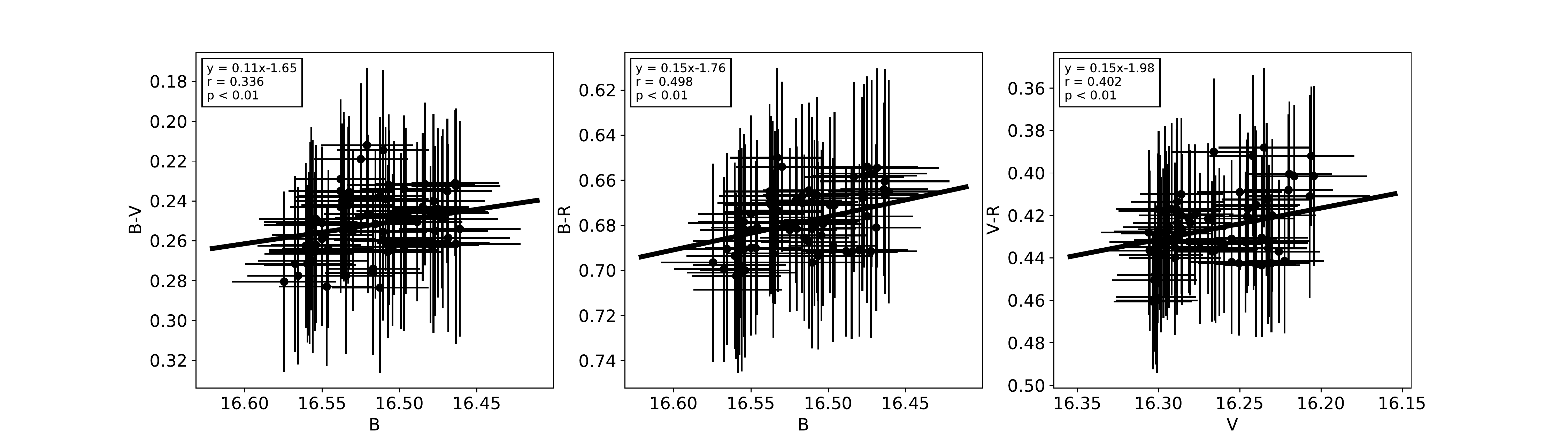}
			\caption{Intraday color-magnitude diagrams of PKS 0735+178 on MJD 59188. \label{intraday_color}}
		\end{center}
	\end{figure}

	In principle, the color index (CI) should be obtained by pairing the simultaneous data points. Hence, our quasi-simultaneous intraday light curves were linearly interpolated to get the intraday CIs. The intraday color behaviors on MJD 59188 were shown in Fig.~\ref{intraday_color}. In order to comprehensively consider the errors in both magnitudes and CIs, we introduced the BCES estimator \citep{Akritas1996}. Only when the absolute value of the correlation coefficient exceeded 0.2 and the statistical significance of the correlation exceeded 99$\%$ (i.e., $p<0.01$), the target was considered to show a confident color-magnitude correlation. It is clear that all the diagrams in Fig.~\ref{intraday_color} show significant BWB trends.
	
\begin{figure}[ht!]
		\begin{center}
			\includegraphics[scale=0.5]{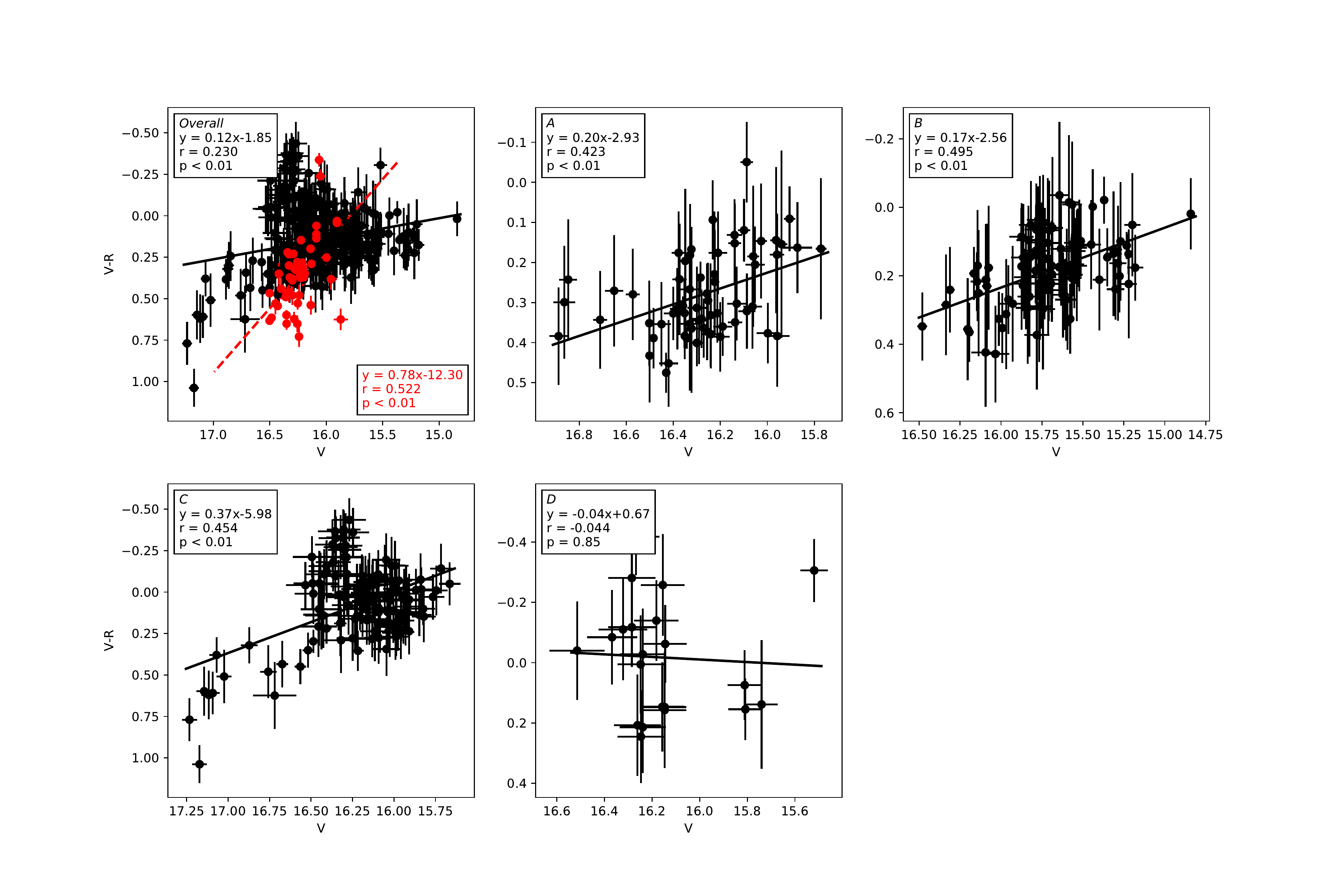}
			\caption{Long-term optical color-magnitude correlations of PKS 0735+178. The panels from top to bottom and left to right represent the overall phase and four individual phases, respectively, and the corresponding formula, correlation coefficient $r$ and $p$ value were given in the top-left corner of each panel. We also fitted those data points without interpolation, which are labeled by the red dots and dashed line in the top-left panel, and the specific parameters were given by the red legends.  \label{optical_color}}
		\end{center}
	\end{figure}

	However, since our long-term optical data were collected from multiple datasets, we can only pair two data points on the same night. Even so, all the obtained CIs clustered in the period of MJD 55000-55400. Thus, we interpolated the long-term light curves with the damped random walk (DRW) method, which was proven to model quasar light curve behavior quite well on time scales of months to years \citep{Gaskell1987, MacLeod2010, Zu2013}. More details can be found in Appendix~\ref{appendix}. We interpolated the $R$-band light curve to obtain the simultaneous data points with those of the $V$ band and calculated the CIs. For comparison, the original CIs without interpolation were also calculated. The variation of CIs was presented in Fig.~\ref{LTLC}.	The long-term color trends of PKS 0735+178 were presented in Fig.~\ref{optical_color}, including the overall trend and four individual trends\footnote{Due to the small sample of data, we did not give the diagram of phase $E$.}.
 	
	The BWB behaviors were exhibited on the overall phase, phases $A$, $B$ and $C$. Such behaviors were also proven by the color-magnitude correlation without interpolation (red dots and dashed line in the top-left panel). Our results are consistent with the common color trend of BL Lac objects \citep{Fan2000}. We noticed that the BWB trend exhibited in phase $B$, which contained a significant flare component, was stronger than that of phase $A$. Several studies have demonstrated that the BWB trend of some blazars will be enhanced during active states, e.g., S5 0716+714 \citep{Wu2007}, PG 1553+113 \citep{Agarwal2021} and AO 0235+164 \citep{Wang2020}. As a comparison, the phase $A$ that contained only one incomplete flare presented a mild BWB trend, and an achromatic trend was exhibited in phase $D$ that lacked a flare component. 
	
\begin{figure}[ht!]
		\begin{center}
			\includegraphics[scale=0.5]{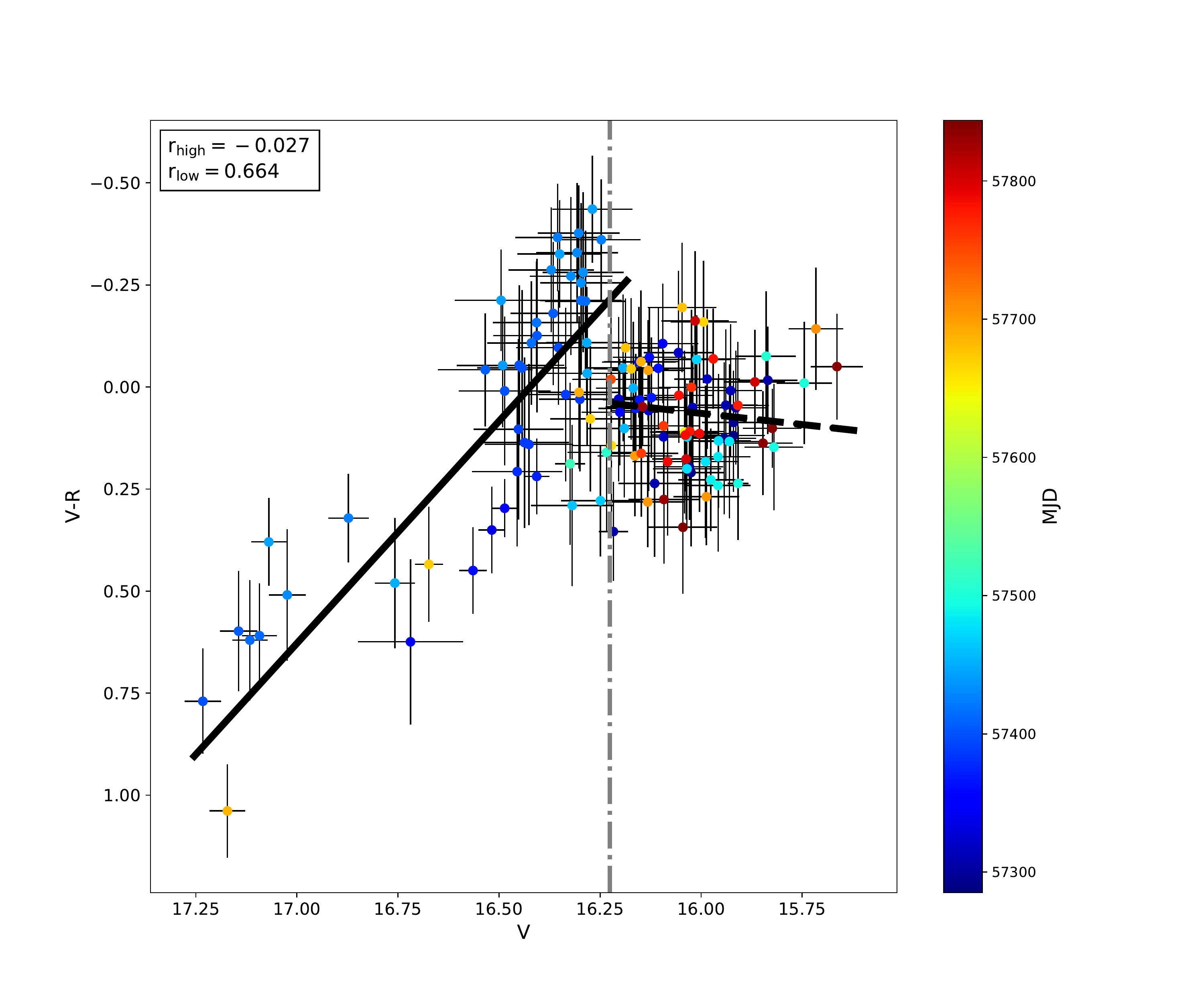}
			\caption{The fittings of the low and high states in phase $C$ divided by the vertical dash-dotted line. The solid and dashed lines represent the fittings of the low and high states, respectively. The corresponding Spearman correlation coefficients are given in the upper left corner. \label{phaseCo}}
		\end{center}
	\end{figure}

	The most complex color behavior was presented in phase $C$. From visual inspection, two distinct components were found in the color-magnitude correlation, i.e., a strong BWB trend and an achromatic one. To further investigate the color behavior, we divided the variations in phase $C$ into low and high states by the mean magnitude ($\sim 16.23$ mags) of this period, as shown in Fig.~\ref{phaseCo}. Particularly, the low state corresponds to the historical low state mentioned above. PKS 0735+178 showed no trend in the high state and a strong BWB trend in the low state. One can find that the correlation coefficient of the low state reached 0.664, which means the most significant BWB trend was observed in the historical low state.

\begin{figure}[ht!]
		\begin{center}
			\includegraphics[scale=0.32]{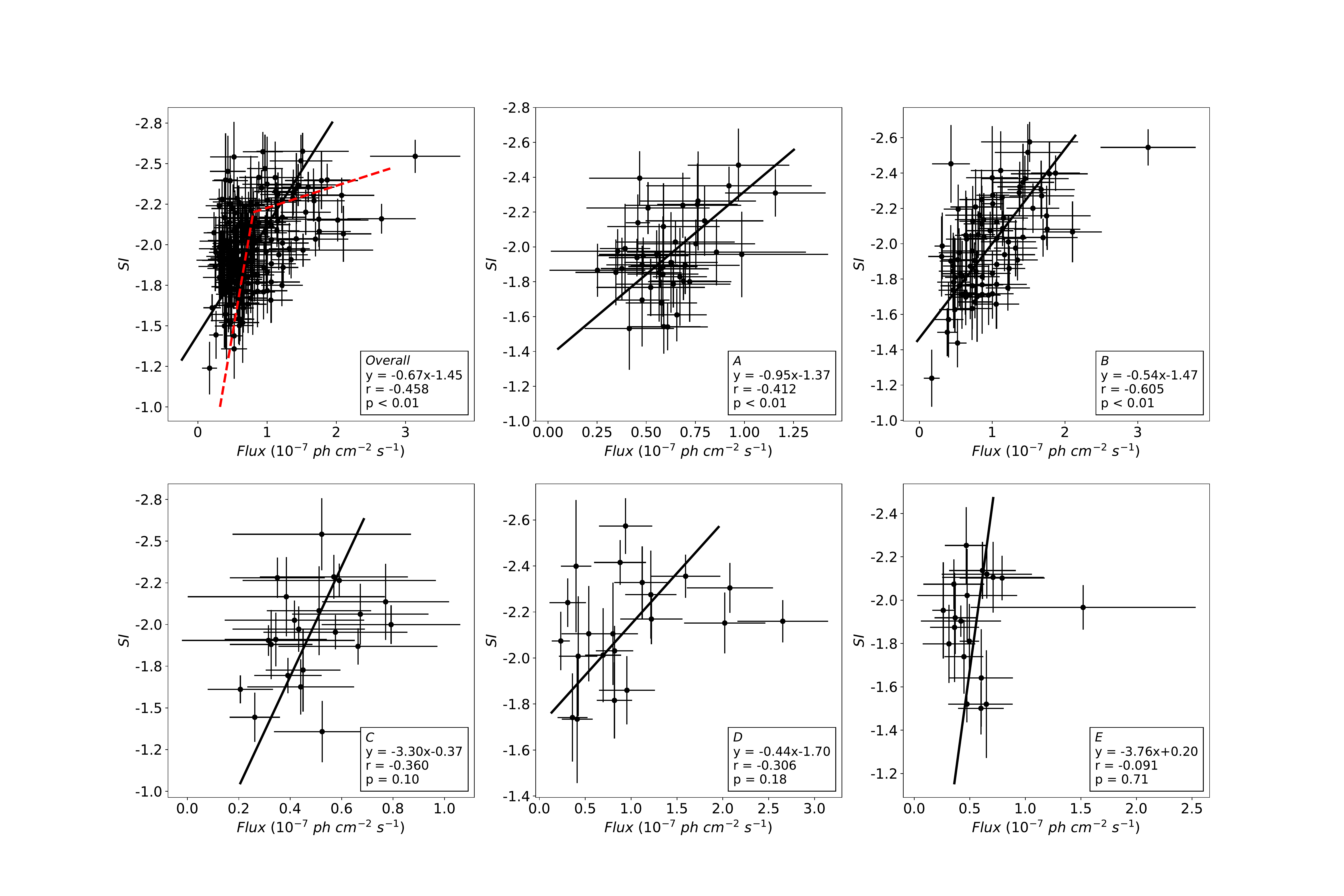}
			\caption{Long-term $\gamma$-ray SI-flux correlations of PKS 0735+178. The panels from top to bottom and from left to right represent the overall phase and five individual phases, respectively. The change of the spectral behavior from a HWB trend to a saturation in the overall phase is labeled by the red dashed lines. \label{gamma_color}}
		\end{center}
	\end{figure}

	The $\gamma$-ray spectral behaviors were presented in Fig.~\ref{gamma_color}. In the past decade, PKS 0735+178 showed a significant HWB trend, which is consistent to the BWB trend in optical bands. Moreover, the overall HWB trend tends to saturate toward higher flux (see the red dashed lines in the first panel). This is a common phenomenon in the X-ray and $\gamma$-ray bands \citep[e.g.,][]{Xue2005,Giebels2007,Weng2020,Acciari2021}, which suggests a more efficient particle acceleration mechanism in the high state \citep{Giebels2007}. As for the individual trends, two significant HWB behaviors appeared in phases $A$ and $B$, and two low significant HWB behavior appeared in phases $C$ and $D$. We noticed that the HWB trend of phase $B$ with flare components was obviously stronger than those of phases $A$, $C$ and $D$. Such phenomenon is consistent with the situations in the optical bands. The flare in phase $D$ was the shortest one and most of the data points were concentrated in the highest state, so the exhibited HWB trend was not so significant. The spectral hardening during the flare state indicates the increasing of detected high energy photons, which is expected if the inverse Compton peak shifts to higher energy \citep{Shah2019,Shah2021}. Moreover, a SWB trend was found with a low significance level in phase $E$. 
	
	\subsection{Interpretation and Discussions on Color Behavior}
	Since the lack of a radiative efficiently accretion disk, the color behavior of BL Lac objects were always interpreted by the jet emission model \citep{Marscher1985}. In the frame of the leptonic model of blazars, the electrons are accelerated near the root of the relativistic jet and the magnetic field is compressed therein. This process leads to the observed flux and spectral variabilities \citep{Marscher1985,Qian1991}. Additionally, the weak intraday BWB behavior also could be explained by the superposition of many distinct new variable components \citep{Gaur2015}.
	
	Furthermore, the significant BWB trend in the historical low state suggests that the acceleration process in flare and low activity states may have different origins.	In principle, we should consider carefully the contamination from the host galaxy in such a low state. However, the brightness of the host galaxy ($>20.8$ mag, \cite{Scarpa2000}) was too faint with respect to the nucleus, and therefore it was unlikely to affect the color behavior. \cite{2013ApJ...765..122Y} found that the shock acceleration is dominant in the low state, while stochastic turbulence acceleration is dominant in the flare state. 
	
	According to the scenario proposed by \cite{Virtanen2005}, the electrons are accelerated at shock front and move along the jet. The shock acceleration will continuously affect electrons, both in the low and high states. However, the turbulence appears rather strong in the downstream jet, and the stochastic turbulence can have considerable contributions to the acceleration process of electrons \citep{Virtanen2005, Marscher2013}. The shock-accelerated electrons are efficiently re-accelerated in the downstream jet through a stochastic turbulence, and the increasing radiation contributes to the total emission and the blazar will turn to a flare state \citep{2013ApJ...765..122Y, Feng2020}. Moreover, the high-energy $\gamma$-ray photons were also thought to be originated in the downstream region \citep{Jorstad2001}. Both shock acceleration and stochastic turbulence acceleration lead to variations in the magnetic field inside the jet, which further change the brightness of the blazar \citep{Kirk1998}. If the inter-band amplitude difference in the low state is larger than that of the flare state, the corresponding color behavior will be more significant \citep{Dai2011}.
	
	Based on a review of previous studies, \cite{Ciprini2007} proposed an achromatic trend of PKS 0735+178. Unfortunately, they lacked the variation data during the 2001 outburst. So they weren’t able to check the spectral behavior. Thus, their data suggested a rather constant SI and presented an achromatic trend. This is not in conflict with our conclusions. A similar achromatic trend caused by the incomplete data was also reported by \cite{Gu2006}. The RWB trend proposed by \cite{Sandrinelli2014} needs an in-depth discussion. We found that their data also did not contain any complete flare. In our analysis, such properties should be the features of an achromatic or weak BWB trend rather than a clear RWB trend. We noticed that their variation analysis involved the infrared $J$-band. \cite{Isler2017} mentioned that the color behavior depended on the ratio of the contributions from the accretion disk and relativistic jet, i.e., the ratio of thermal emission to non-thermal emission. In the infrared band, the thermal emission is thought to partly come from the dusty molecular torus that reprocesses the photons from the broad line region and the accretion disk into the infrared region \citep{Antonucci1993,Perlman2008}. This torus component increases the proportion of the thermal emission and causes the source to exhibit a RWB trend. Moreover, \cite{Safna2020} also showed that the RWB seems a common trend among blazars when an infrared band was involved. 
	
\begin{figure}[ht!]
		\begin{center}
			\includegraphics[scale=0.5]{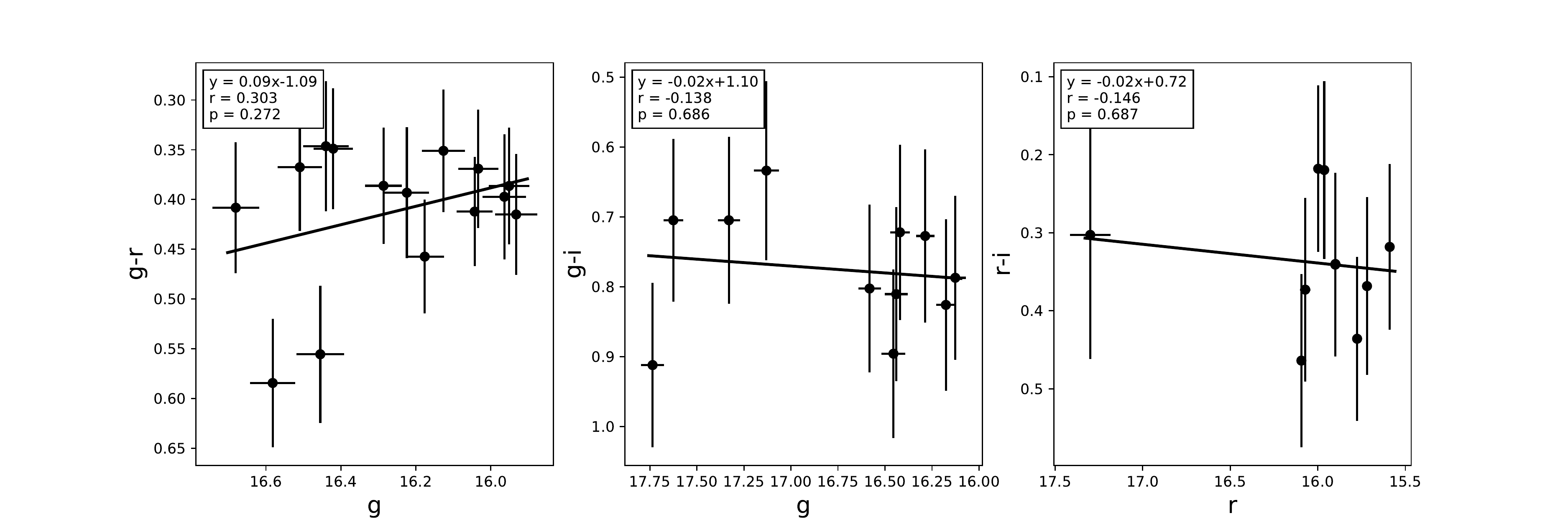}
			\caption{Reconstructed color-magnitude correlations. The nightly averaged magnitudes were used. \label{Yuan2021}}
		\end{center}
	\end{figure}

	The two opposite color trends reported by \cite{2021PASP..133g4101Y} seem to interfere by the clustering of the data points (see Fig. 9 therein). We noticed that they didn't take a nightly average for those data points, which would have affected the final result by superimposing the intraday trend on the long-term trend. To verify our idea, we reconstructed the color-magnitude correlations of their data with our method, as shown in Fig.~\ref{Yuan2021}. Unfortunately, all the three diagrams didn't show any significant correlations, which were mainly caused by the small sample of data points and associated large uncertainties. So we couldn't conclude which color-magnitude correlation exhibited during their observations.
	
	\section{Summary}\label{Sec5}
	In this work, we monitored PKS 0735+178 with the 85cm telescope at Xinglong observatory for three nights, and we also collected 11-years multi-wavelength light curves of this source. The inter-band time lags were calculated through ICCF and simply discussed the correlations between various bands. We analyzed the color and spectral behavior to investigate the variation mechanism. Our conclusions are summarized as follows:
	\begin{itemize}
	\item IDV was observed on MJD 59188. Based on the results of ICCF, the only one non-zero lag of $22.1$d was detected between the $R$ and $\gamma$-ray bands. Too close emitting regions prevent a non-zero lag been detected between two optical bands.
	\item The correlation between the $\gamma$-ray and $R$ bands is much weaker than that between the $V$ and $R$ bands. This is mainly due to the different origins of the optical and $\gamma$-ray emission.
	\item We found that PKS 0735+178 showed a BWB trend in the optical bands on both long-term and intraday timescales, and a HWB trend was found in the $\gamma$-ray band. These color or spectral behaviors will be enhanced during the active states. Such color/spectral behaviors could be naturally explained by the jet emission model. 
	\item The most significant BWB trend was found during the historical low state. It could be explained by the larger difference of the inter-band variation amplitude of the trough state compared to that of the flare state, which may be caused by the acceleration process with different origins. Moreover, the saturation of HWB trend in the high state suggests an efficient particle acceleration mechanism.
	\end{itemize}

	Either as a blazar with special color behaviors or as a potential or even promising neutrino emitter, PKS 0735+178 is an interesting target and can be monitored intensively at multiple electromagnetic wavelengths and, especially, with IceCube or future neutrino telescope (e.g., Cubic Kilometre Neutrino Telescope), in order to disclose the mechanisms and regions responsible for the neutrino and electromagnetic emissions. Improved localization and sensitivity on neutrino detection is required, and a joint electromagnetic-neutrino observations will be a key in identifying flaring neutrino sources.

	\acknowledgements
	The authors thank the anonymous referee for the valuable comments which improved the quality and clarity of the manuscript. This work has been supported by the Chinese National Natural Science Foundation nts 11973017. Data from the Steward Observatory spectropolarimetric monitoring project were used. This program is supported by Fermi Guest Investigator grants NNX08AW56G, NNX09AU10G, NNX12AO93G, and NNX15AU81G. The CRTS survey is supported by the U.S.~National Science Foundation under grants AST-0909182.
	
	\software{IRAF \citep{Tody1986,Tody1993}, BCES \citep{Akritas1996,Nemmen2012}, PyCCF \citep{Peterson1998a,Sun2018}, JAVELIN \citep{Zu2011}.}
	
	\appendix 
	\renewcommand\thefigure{\Alph{section}\arabic{figure}}
	\setcounter{figure}{0}
	\section{DRW Interpolation}\label{appendix}
\begin{figure}[ht!]
		\begin{center}
			\includegraphics[scale=0.5]{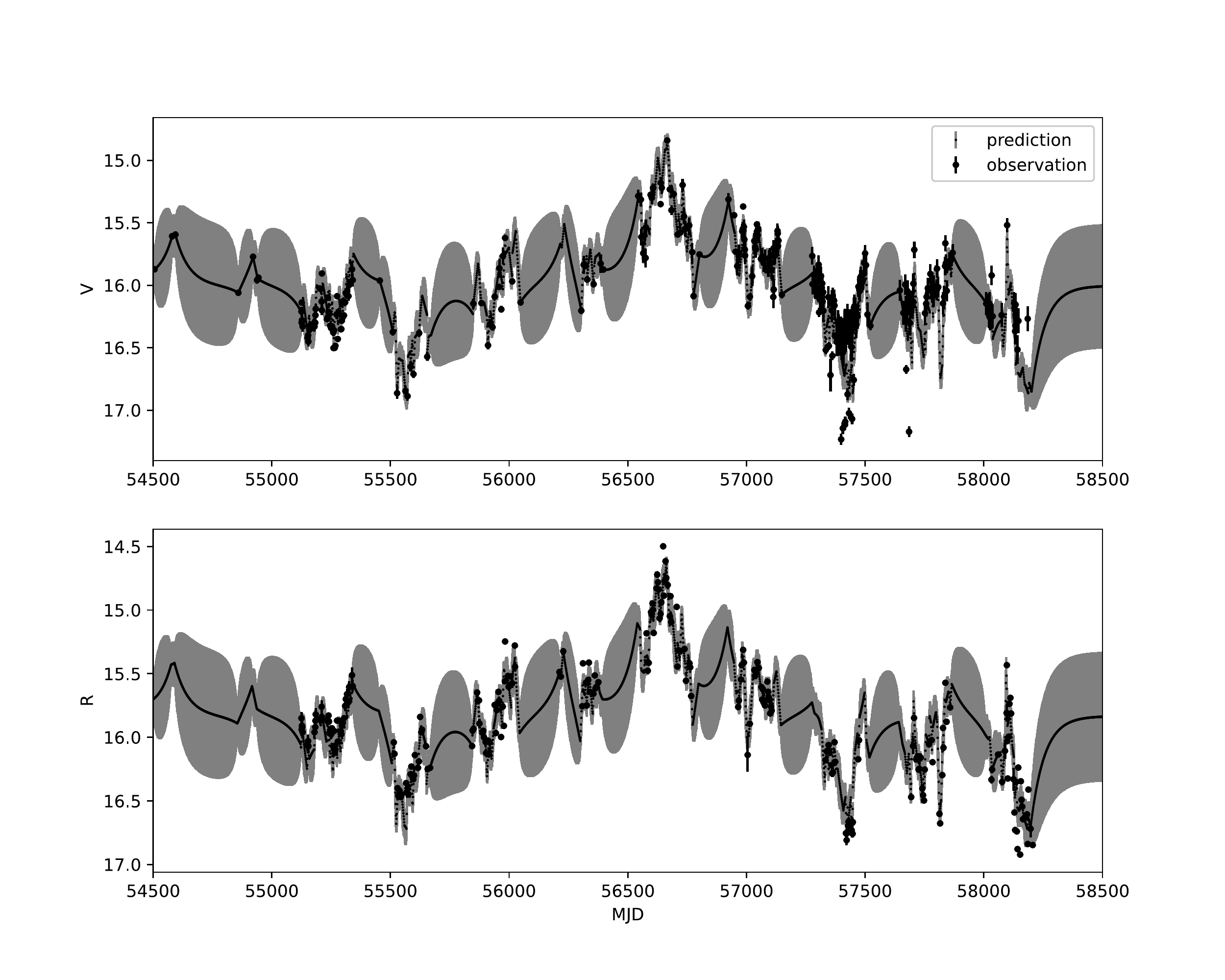}
			\caption{The predicted light curves of $V$ and $R$ bands. The shade region indicates the error bound and the black points indicate the observations.\label{prediction}}
		\end{center}
	\end{figure}

	The covariance function of a DRW has an exponential form
	\begin{equation*}
	S(\Delta t) = \sigma_{DRW}^{2}\exp(-|\Delta t/\tau_{DRW}|)
	\end{equation*}
	where $\Delta t$ is the time interval between two epochs, $\sigma_{DRW}$ is the DRW amplitude, and $\tau_{DRW}$ is the DRW timescale. The DRW interpolation was implemented through JAVELIN\footnote{\url{https://github.com/nye17/javelin}} algorithm \citep{Zu2011}. First we fitted the $R$-band light curve to constrain the $\sigma_{DRW}$ and $\tau_{DRW}$. The $V$-band light curve was considered as a shifted, smoothed and scaled version of the $R$-band light curve and shared the same $\tau_{DRW}$ and $\sigma_{DRW}$. In this process, four additional parameters are added, i.e., the time lag, the kernel width, the transfer function amplitude, and the ratio between the two bands. Through the Markov chain Monte Carlo (MCMC) method, we can obtain the best fitting parameters and use them to predict the two light curves. Furthermore, we restricted the time lags between $V$ and $R$ bands in $[-10, 10]$ days, and the DRW timescale have been restricted in $[100, 300]$ days, which is found for a larger sample of quasars from the Sloan Digital Sky Survey \citep{MacLeod2010}. The final predictions of the two light curves were presented in Fig.~\ref{prediction}.
	
	To check whether the interpolated light curve preserves the properties of the original light curve, its time-domain power density spectrum (PSD), i.e., structure function (SF), was calculated. The SF describes the growth of variability with time and is widely used in the analysis of AGN variabilities \citep[e.g.,][]{Ulrich1997,Abdo2010,Ackermann2011}. Compared to the PSD in the frequency domain, the SF is less subject to sampling problems in the presence of very irregular time series \citep{Simonetti1985,Paltani1997}. One of the most common form of SF is defined as
	\begin{equation*}
		SF(\tau)=\frac{1}{N(\tau)} \sum_{i=1}\{[m(t_{i})-m(t_{i}+\tau)]^{2}-[\sigma_{err}(t_i)^2+\sigma_{err}(t_i+\tau)^2]\},
	\end{equation*}
	where $\tau$ is the time difference between the data pair, $m(t_i)$ and $\sigma_{err}(t_i)$ are the magnitude and associated uncertainty at time $t_i$, respectively \citep{Bauer2009}. The uncertainty of SF is defined as
	\begin{equation*}
		\sigma_{SF}(\tau)  = \frac{1}{N_{{i}} \cdot SF} \sqrt{\sum_{i=1}\left\{[m(t_i+\tau)-m(t_i)]^{2} \cdot\left[\sigma_{{err}}^{2}(t_i+\tau)+\sigma_{{err}}^{2}(t_i)\right]\right\}},
	\end{equation*}
	where $N_{i}$ is the number of data point pairs in the bin. In our settings, a linear bin $\tau=4$ days was adopted, and the time scales $\tau>200$ days was discarded.

\begin{figure}[ht!]
		\begin{center}
			\includegraphics[scale=0.3]{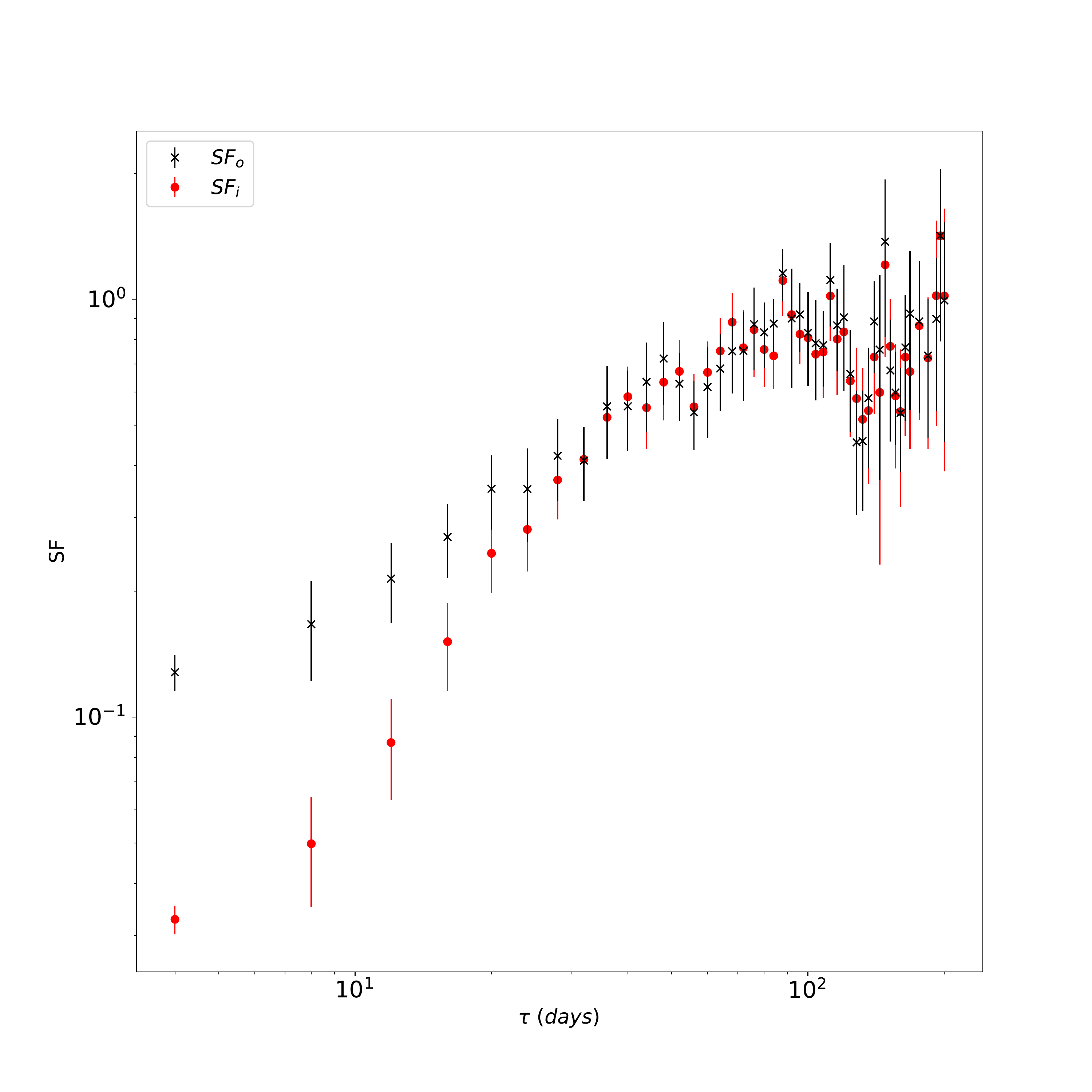}
			\caption{The SFs of the $R$-band interpolated light curve and original light curve, which are denoted by the black crosses and red dots, respectively. \label{compare_sf}}
		\end{center}
	\end{figure}

	We calculated the SFs of the $R$-band interpolated light curve ($SF_i$) and original light curve ($SF_o$), and the comparison between the two SFs were presented in Fig.~\ref{compare_sf}. Notice that our CIs were obtained through the interpolated $R$-band light curve and the original $V$-band light curve, and thus the comparison between the two SFs in the $V$ band was not given. Except for the deviation in the low $\tau$ case ($\tau<32$ days), the two SFs are well consistent with each other, which proves that the interpolation preserves the properties of the original light curve. One possible reason for the low-$\tau$ deviation is the poor fit of the DRW model to the short timescale light curves.

\begin{figure}[ht!]
	\begin{center}
		\includegraphics[scale=0.3]{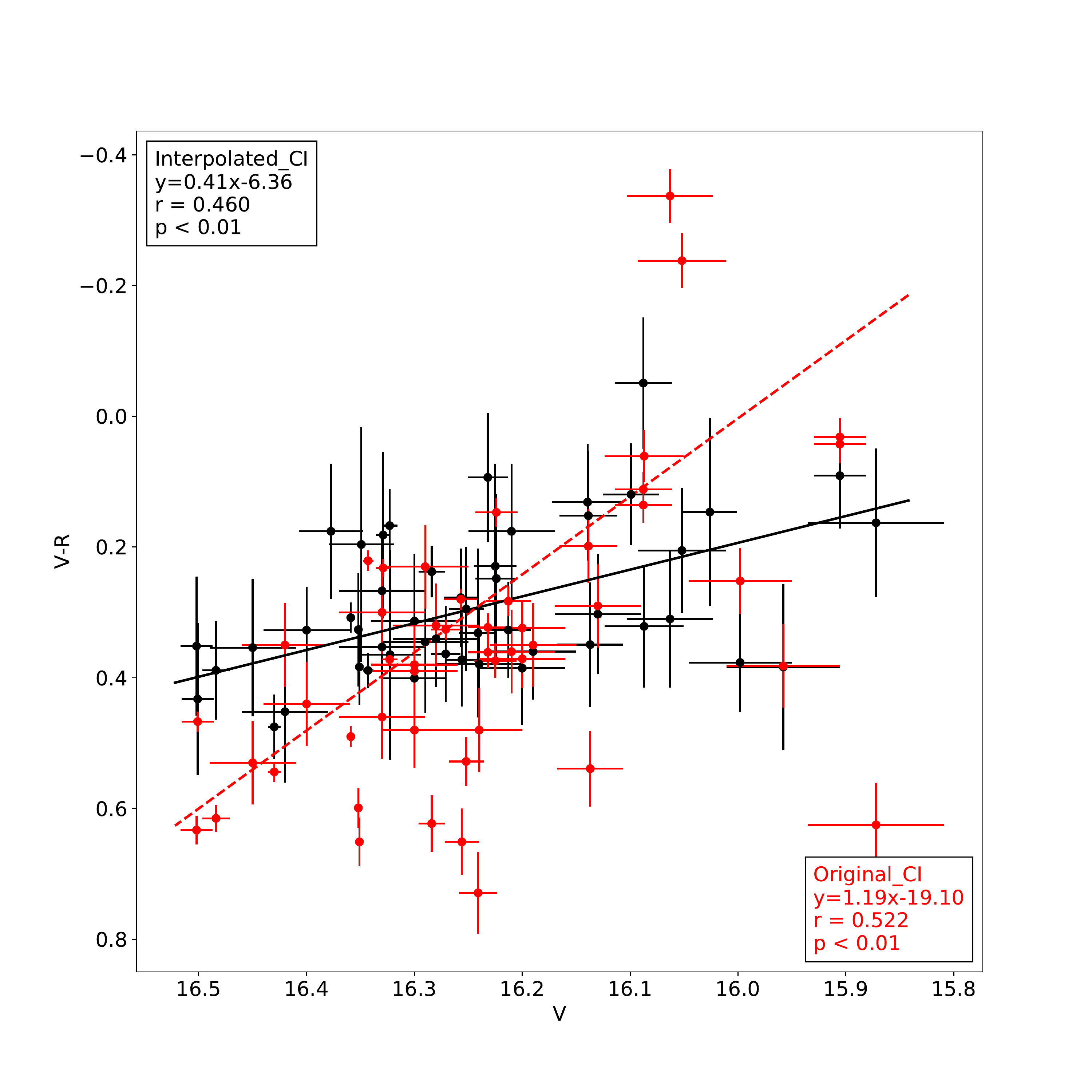}
		\caption{The comparison between the CIs obtained with and without interpolation, which were denoted by the black and red dots, respectively. The corresponding formula, correlation coefficient $r$ and $p$ value are given. \label{compare_CIs}}
	\end{center}
	\end{figure}

	As stated in Sec.~\ref{Sec4.1}, we obtained the CIs without interpolation during MJD 55000-55400, and a comparison between the two CIs would also be a good visualization to the impacts of interpolation. As shown in Fig.~\ref{compare_CIs}, the CIs obtained with and without interpolation are presented. The time baseline of the two CIs was restricted to the same period, i.e., MJD 55000-55400. It's clear that the color-magnitude correlation with interpolation is only sightly weaker than that without interpolation. It suggests that DRW interpolation does not produce a spurious correlation. Unfortunately, we don't have more data to extend the time baseline to assess its performance on longer timescale.

	\section{ICCF of $R$-$J$ bands}\label{infrared}
\begin{figure}[ht!]
	\begin{center}
		\includegraphics[scale=0.5]{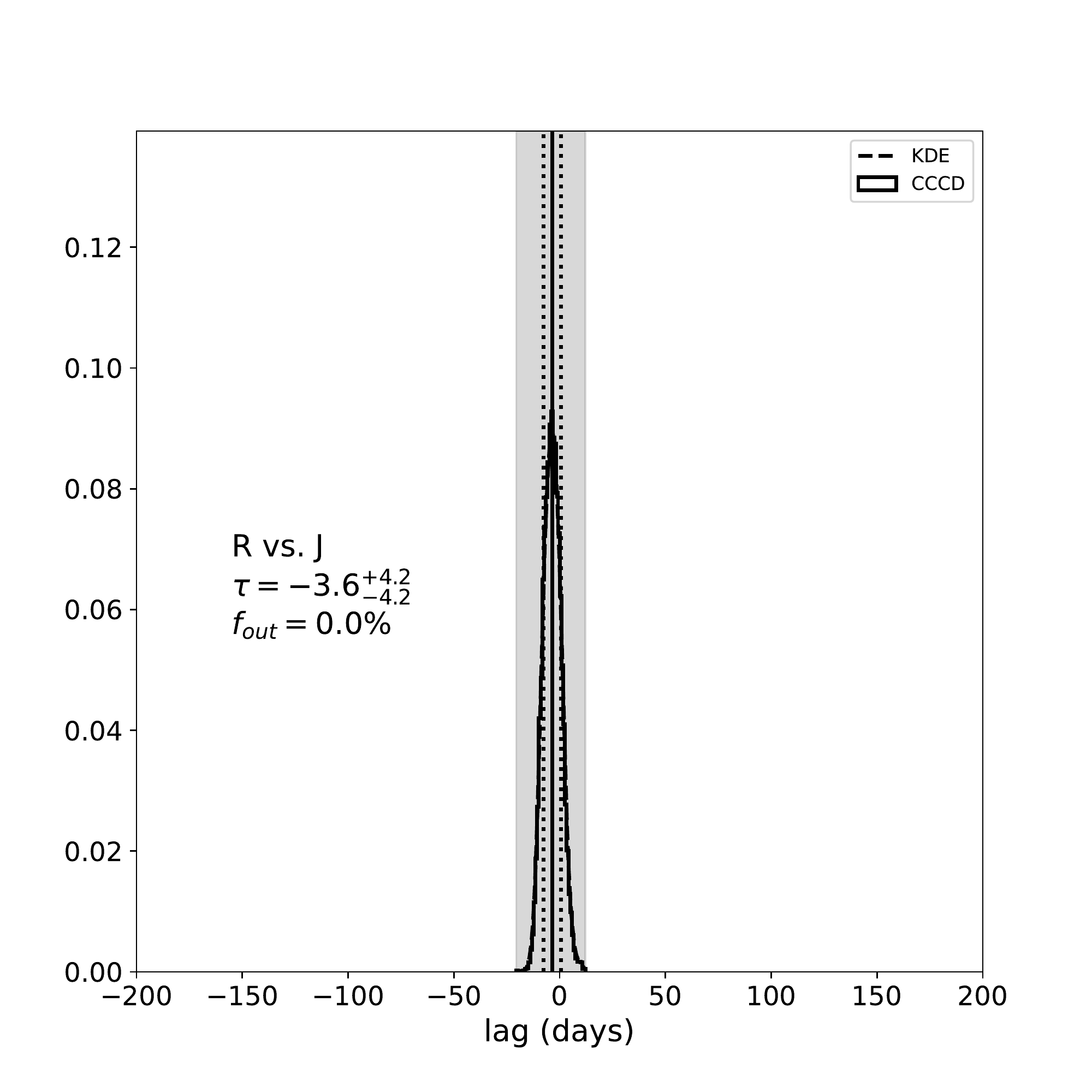}
		\caption{The same as Fig.~\ref{lag}, but for $R-J$ bands. \label{RJ}}
	\end{center}
	\end{figure}

	We used the raw data from \cite{Sandrinelli2014} to implement the ICCF. We restricted the search range of time lags to $[-200, 200]$ days and 10000 MC realizations were applied. The final result showed a $0\%$ failure rate, $0\%$ removal efficiency and a maximum correlation coefficient $\sim 0.8$, which indicate a reliable measurement, and a $-3.6^{+4.2}_{-4.2}$ days lag was detected (see Fig.~\ref{RJ}).

\newpage
\bibliographystyle{aasjournal}
\bibliography{reference}

\begin{thebibliography}{}
\expandafter\ifx\csname natexlab\endcsname\relax\def\natexlab#1{#1}\fi
\providecommand{\url}[1]{\href{#1}{#1}}

\bibitem[{Abdo {et~al.}(2010)Abdo, Ackermann, Ajello, Antolini, Baldini,
  Ballet, Barbiellini, Bastieri, Bechtol, Bellazzini, Berenji, Blandford,
  Bloom, Bonamente, Borgland, Bouvier, Bregeon, Brez, Brigida, Bruel, Buehler,
  Burnett, Buson, Caliandro, Cameron, Caraveo, Carrigan, Casandjian, Cavazzuti,
  Cecchi, {\c{C}}elik, Chekhtman, Cheung, Chiang, Ciprini, Claus, Cohen-Tanugi,
  Cominsky, Conrad, Costamante, Cutini, Dermer, de~Angelis, de~Palma, Silva,
  Drell, Dubois, Dumora, Farnier, Favuzzi, Fegan, Focke, Fortin, Frailis,
  Fukazawa, Funk, Fusco, Gargano, Gasparrini, Gehrels, Germani, Giebels,
  Giglietto, Giommi, Giordano, Glanzman, Godfrey, Grenier, Grondin, Grove,
  Guiriec, Hadasch, Hayashida, Hays, Healey, Horan, Hughes, Itoh,
  J{\'o}hannesson, Johnson, Johnson, Kamae, Katagiri, Kataoka, Kawai,
  Kn{\"o}dlseder, Kuss, Lande, Larsson, Latronico, Lemoine-Goumard, Longo,
  Loparco, Lott, Lovellette, Lubrano, Madejski, Makeev, Massaro, Mazziotta,
  McEnery, Michelson, Mitthumsiri, Mizuno, Moiseev, Monte, Monzani, Morselli,
  Moskalenko, Mueller, Murgia, Nolan, Norris, Nuss, Ohno, Ohsugi, Omodei,
  Orlando, Ormes, Ozaki, Panetta, Parent, Pelassa, Pepe, Pesce-Rollins, Piron,
  Porter, Rain{\`o}, Rando, Razzano, Reimer, Reimer, Ritz, Rodriguez, Romani,
  Roth, Ryde, Sadrozinski, Sander, Scargle, Sgr{\`o}, Shaw, Smith, Spandre,
  Spinelli, Starck, Strickman, Suson, Takahashi, Takahashi, Tanaka, Thayer,
  Thayer, Thompson, Tibaldo, Torres, Tosti, Tramacere, Uchiyama, Usher,
  Vasileiou, Vilchez, Vitale, Waite, Wallace, Wang, Winer, Wood, Yang, Ylinen,
  \& Ziegler}]{Abdo2010}
Abdo, A.~A., Ackermann, M., Ajello, M., {et~al.} 2010, \apj, 722, 520.
\newblock \url{https://ui.adsabs.harvard.edu/abs/2010ApJ...722..520A}

\bibitem[{Acciari {et~al.}(2021)Acciari, Ansoldi, Antonelli, Asano, Babi{\'c},
  Banerjee, Baquero, de~Almeida, Barrio, Becerra~Gonz{\'a}lez, Bednarek,
  Bellizzi, Bernardini, Bernardos, Berti, Besenrieder, Bhattacharyya,
  Bigongiari, Blanch, Bonnoli, Bo{\v{s}}njak, Busetto, Carosi, Ceribella,
  Cerruti, Chai, Chilingarian, Cikota, Colak, Colombo, Contreras, Cortina,
  Covino, D'Amico, D'Elia, Da~Vela, Dazzi, De~Angelis, De~Lotto, Delfino,
  Delgado, Delgado~Mendez, Depaoli, Di~Girolamo, Di~Pierro, Di~Venere,
  Do~Souto~Espi{\~n}eira, Dominis~Prester, Donini, Doro, Fallah~Ramazani,
  Fattorini, Ferrara, Foffano, Fonseca, Font, Fruck, Fukami,
  Garc{\'\i}a~L{\'o}pez, Garczarczyk, Gasparyan, Gaug, Giglietto, Giordano,
  Gliwny, Godinovi{\'c}, Green, Green, Hadasch, Hahn, Heckmann, Herrera, Hoang,
  Hrupec, H{\"u}tten, Inada, Inoue, Ishio, Iwamura, Jormanainen, Jouvin,
  Kajiwara, Karjalainen, Kerszberg, Kobayashi, Kubo, Kushida, Lamastra, Lelas,
  Leone, Lindfors, Lombardi, Longo, L{\'o}pez, L{\'o}pez-Coto,
  L{\'o}pez-Oramas, Loporchio, Machado~de Oliveira~Fraga, Maggio, Majumdar,
  Makariev, Mallamaci, Maneva, Manganaro, Maraschi, Mariotti, Mart{\'\i}nez,
  Mazin, Mender, Mi{\'c}anovi{\'c}, Miceli, Miener, Minev, Miranda, Mirzoyan,
  Molina, Moralejo, Morcuende, Moreno, Moretti, Munar-Adrover, Neustroev,
  Nigro, Nilsson, Ninci, Nishijima, Noda, Nozaki, Ohtani, Oka, Otero-Santos,
  Palatiello, Paneque, Paoletti, Paredes, Pavleti{\'c}, Pe{\~n}il, Perennes,
  Persic, Prada~Moroni, Prandini, Priyadarshi, Puljak, Rhode, Rib{\'o}, Rico,
  Righi, Rugliancich, Saha, Sahakyan, Saito, Sakurai, Satalecka, Schleicher,
  Schmidt, Schweizer, Sitarek, {\v{S}}nidari{\'c}, Sobczynska, Spolon,
  Stamerra, Strom, Strzys, Suda, Suri{\'c}, Takahashi, Tavecchio, Temnikov,
  Terzi{\'c}, Teshima, Torres-Alb{\`a}, Tosti, Truzzi, van Scherpenberg, Vanzo,
  Vazquez~Acosta, Ventura, Verguilov, Vigorito, Vitale, Vovk, Will, Zari{\'c},
  Arbet-Engels, Baack, Balbo, Beck, Biederbeck, Biland, Bretz, Bruegge, Buss,
  Dorner, Elsaesser, Hildebrand, Iotov, Klinger, Mannheim, Neise, Neronov,
  Noethe, Paravac, Rhode, Schleicher, Sliusar, Theissen, Walter, Valverde,
  Horan, Giroletti, Perri, Verrecchia, Leto, Sadun, Moody, Joner, Marscher,
  Jorstad, L{\"a}hteenm{\"a}ki, Tornikoski, Ramakrishnan, J{\"a}rvel{\"a},
  Vera, Righini, \& Lien}]{Acciari2021}
Acciari, V.~A., Ansoldi, S., Antonelli, L.~A., {et~al.} 2021, \mnras, 504, 1427

\bibitem[{Ackermann {et~al.}(2011)Ackermann, Ajello, Allafort, Antolini,
  Atwood, Axelsson, Baldini, Ballet, Barbiellini, Bastieri, Bechtol,
  Bellazzini, Berenji, Blandford, Bloom, Bonamente, Borgland, Bottacini,
  Bouvier, Bregeon, Brigida, Bruel, Buehler, Burnett, Buson, Caliandro,
  Cameron, Caraveo, Casandjian, Cavazzuti, Cecchi, Charles, Cheung, Chiang,
  Ciprini, Claus, Cohen-Tanugi, Conrad, Costamante, Cutini, de~Angelis,
  de~Palma, Dermer, Digel, Silva, Drell, Dubois, Escande, Favuzzi, Fegan,
  Ferrara, Finke, Focke, Fortin, Frailis, Fukazawa, Funk, Fusco, Gargano,
  Gasparrini, Gehrels, Germani, Giebels, Giglietto, Giommi, Giordano,
  Giroletti, Glanzman, Godfrey, Grenier, Grove, Guiriec, Gustafsson, Hadasch,
  Hayashida, Hays, Healey, Horan, Hou, Hughes, Iafrate, J{\'o}hannesson,
  Johnson, Johnson, Kamae, Katagiri, Kataoka, Kn{\"o}dlseder, Kuss, Lande,
  Larsson, Latronico, Longo, Loparco, Lott, Lovellette, Lubrano, Madejski,
  Mazziotta, McConville, McEnery, Michelson, Mitthumsiri, Mizuno, Moiseev,
  Monte, Monzani, Moretti, Morselli, Moskalenko, Murgia, Nakamori,
  Naumann-Godo, Nolan, Norris, Nuss, Ohno, Ohsugi, Okumura, Omodei, Orienti,
  Orlando, Ormes, Ozaki, Paneque, Parent, Pesce-Rollins, Pierbattista,
  Piranomonte, Piron, Pivato, Porter, Rain{\`o}, Rando, Razzano, Razzaque,
  Reimer, Reimer, Ritz, Rochester, Romani, Roth, Sanchez, Sbarra, Scargle,
  Schalk, Sgr{\`o}, Shaw, Siskind, Spandre, Spinelli, Strong, Suson, Tajima,
  Takahashi, Takahashi, Tanaka, Thayer, Thayer, Thompson, Tibaldo, Tinivella,
  Torres, Tosti, Troja, Uchiyama, Vandenbroucke, Vasileiou, Vianello, Vitale,
  Waite, Wallace, Wang, Winer, Wood, Wood, \& Zimmer}]{Ackermann2011}
Ackermann, M., Ajello, M., Allafort, A., {et~al.} 2011, \apj, 743, 171.
\newblock \url{https://ui.adsabs.harvard.edu/abs/2011ApJ...743..171A}

\bibitem[{Agarwal {et~al.}(2021)Agarwal, Mihov, Andruchow, Cellone, Anupama,
  Agrawal, Zola, Slavcheva-Mihova, {\"O}zd{\"o}nmez, Ege, Raj, Mammana,
  Zibecchi, \& Fern{\'a}ndez-Laj{\'u}s}]{Agarwal2021}
Agarwal, A., Mihov, B., Andruchow, I., {et~al.} 2021, \aap, 645, A137

\bibitem[{Akritas \& Bershady(1996)}]{Akritas1996}
Akritas, M.~G., \& Bershady, M.~A. 1996, \apj, 470, 706.
\newblock \url{https://ui.adsabs.harvard.edu/abs/1996ApJ...470..706A}

\bibitem[{Antonucci(1993)}]{Antonucci1993}
Antonucci, R. 1993, \araa, 31, 473

\bibitem[{Bauer {et~al.}(2009)Bauer, Baltay, Coppi, Ellman, Jerke, Rabinowitz,
  \& Scalzo}]{Bauer2009}
Bauer, A., Baltay, C., Coppi, P., {et~al.} 2009, \apj, 696, 1241.
\newblock \url{https://ui.adsabs.harvard.edu/abs/2009ApJ...696.1241B}

\bibitem[{{B{\"o}ttcher}(2007)}]{Bottcher2007}
{B{\"o}ttcher}, M. 2007, \apss, 309, 95

\bibitem[{{Bregman} {et~al.}(1984){Bregman}, {Glassgold}, {Huggins}, {Aller},
  {Aller}, {Hodge}, {Rieke}, {Lebofsky}, {Pollock}, {Pica}, {Leacock}, {Smith},
  {Webb}, {Balonek}, {Dent}, {O'Dea}, {Ku}, {Schwartz}, {Miller}, {Rudy}, \&
  {Levan}}]{Bregman1984}
{Bregman}, J.~N., {Glassgold}, A.~E., {Huggins}, P.~J., {et~al.} 1984, \apj,
  276, 454

\bibitem[{Carini {et~al.}(2011)Carini, Walters, \& Hopper}]{Carini2011}
Carini, M.~T., Walters, R., \& Hopper, L. 2011, \aj, 141, 49

\bibitem[{{Carswell} {et~al.}(1974){Carswell}, {Strittmatter}, {Williams},
  {Kinman}, \& {Serkowski}}]{Carswell1974}
{Carswell}, R.~F., {Strittmatter}, P.~A., {Williams}, R.~E., {Kinman}, T.~D.,
  \& {Serkowski}, K. 1974, \apjl, 190, L101

\bibitem[{{Chatterjee} {et~al.}(2013){Chatterjee}, {Fossati}, {Urry}, {Bailyn},
  {Maraschi}, {Buxton}, {Bonning}, {Isler}, \& {Coppi}}]{2013ApJ...763L..11C}
{Chatterjee}, R., {Fossati}, G., {Urry}, C.~M., {et~al.} 2013, \apjl, 763, L11

\bibitem[{{Ciprini} {et~al.}(2007){Ciprini}, {Takalo}, {Tosti}, {Raiteri},
  {Fiorucci}, {Villata}, {Nucciarelli}, {Lanteri}, {Nilsson}, \&
  {Ros}}]{Ciprini2007}
{Ciprini}, S., {Takalo}, L.~O., {Tosti}, G., {et~al.} 2007, \aap, 467, 465

\bibitem[{{Cohen} {et~al.}(2014){Cohen}, {Romani}, {Filippenko}, {Cenko},
  {Lott}, {Zheng}, \& {Li}}]{2014ApJ...797..137C}
{Cohen}, D.~P., {Romani}, R.~W., {Filippenko}, A.~V., {et~al.} 2014, \apj, 797,
  137

\bibitem[{Dai {et~al.}(2011)Dai, Wu, Zhu, Zhou, \& Ma}]{Dai2011}
Dai, Y., Wu, J., Zhu, Z.-H., Zhou, X., \& Ma, J. 2011, \aj, 141, 65

\bibitem[{de~Diego(2014)}]{Diego2014}
de~Diego, J.~A. 2014, \aj, 148, 93

\bibitem[{de~Diego {et~al.}(1998)de~Diego, Dultzin-Hacyan, Ram{\'\i}rez, \&
  Ben{\'\i}tez}]{Diego1998}
de~Diego, J.~A., Dultzin-Hacyan, D., Ram{\'\i}rez, A., \& Ben{\'\i}tez, E.
  1998, \apj, 501, 69

\bibitem[{Drake {et~al.}(2009)Drake, Djorgovski, Mahabal, Beshore, Larson,
  Graham, Williams, Christensen, Catelan, Boattini, Gibbs, Hill, \&
  Kowalski}]{Drake2009}
Drake, A.~J., Djorgovski, S.~G., Mahabal, A., {et~al.} 2009, \apj, 696, 870

\bibitem[{{Fan} \& {Lin}(2000)}]{Fan2000}
{Fan}, J.~H., \& {Lin}, R.~G. 2000, \apj, 537, 101

\bibitem[{Fan {et~al.}(1997)Fan, Xie, Lin, Qin, Li, \& Zhang}]{Fan1997}
Fan, J.~H., Xie, G.~Z., Lin, R.~G., {et~al.} 1997, \aaps, 125, 525

\bibitem[{Feng {et~al.}(2020)Feng, Yang, Yang, Liu, Bai, Li, Zhao, Zhang, Li,
  Xiao, Xin, Xing, Lu, Xu, Wang, Wang, Zhang, Zhang, Lun, \& He}]{Feng2020}
Feng, H.-C., Yang, S., Yang, Z.-X., {et~al.} 2020, \apj, 902, 42

\bibitem[{Gaskell \& Peterson(1987)}]{Gaskell1987}
Gaskell, C.~M., \& Peterson, B.~M. 1987, \apjs, 65, 1

\bibitem[{{Gaur} {et~al.}(2012){Gaur}, {Gupta}, {Strigachev}, {Bachev},
  {Semkov}, {Wiita}, {Peneva}, {Boeva}, {Slavcheva-Mihova}, {Mihov}, {Latev},
  \& {Pandey}}]{Gaur2012a}
{Gaur}, H., {Gupta}, A.~C., {Strigachev}, A., {et~al.} 2012, \mnras, 425, 3002

\bibitem[{Gaur {et~al.}(2015)Gaur, Gupta, Bachev, Strigachev, Semkov,
  B{\"o}ttcher, Wiita, de~Diego, Gu, Guo, Joshi, Mihov, Palma, Peneva,
  Rajasingam, \& Slavcheva-Mihova}]{Gaur2015}
Gaur, H., Gupta, A.~C., Bachev, R., {et~al.} 2015, \mnras, 452, 4263

\bibitem[{Giebels {et~al.}(2007)Giebels, Dubus, \& Kh{\'e}lifi}]{Giebels2007}
Giebels, B., Dubus, G., \& Kh{\'e}lifi, B. 2007, \aap, 462, 29

\bibitem[{{Goyal} {et~al.}(2009){Goyal}, {Gopal-Krishna}, {Anupama}, {Sahu},
  {Sagar}, {Britzen}, {Karouzos}, {Aller}, \& {Aller}}]{Goyal2009}
{Goyal}, A., {Gopal-Krishna}, {Anupama}, G.~C., {et~al.} 2009, \mnras, 399,
  1622

\bibitem[{{Grier} {et~al.}(2017){Grier}, {Trump}, {Shen}, {Horne}, {Kinemuchi},
  {McGreer}, {Starkey}, {Brand t}, {Hall}, {Kochanek}, {Chen}, {Denney},
  {Greene}, {Ho}, {Homayouni}, {I-Hsiu Li}, {Pei}, {Peterson}, {Petitjean},
  {Schneider}, {Sun}, {AlSayyad}, {Bizyaev}, {Brinkmann}, {Brownstein},
  {Bundy}, {Dawson}, {Eftekharzadeh}, {Fernand ez-Trincado}, {Gao},
  {Hutchinson}, {Jia}, {Jiang}, {Oravetz}, {Pan}, {Paris}, {Ponder}, {Peters},
  {Rogerson}, {Simmons}, {Smith}, \& {Wang}}]{Grier2017}
{Grier}, C.~J., {Trump}, J.~R., {Shen}, Y., {et~al.} 2017, \apj, 851, 21

\bibitem[{{Gu} {et~al.}(2006){Gu}, {Lee}, {Pak}, {Yim}, \& {Fletcher}}]{Gu2006}
{Gu}, M.~F., {Lee}, C.~U., {Pak}, S., {Yim}, H.~S., \& {Fletcher}, A.~B. 2006,
  \aap, 450, 39

\bibitem[{{Hartman} {et~al.}(1999){Hartman}, {Bertsch}, {Bloom}, {Chen},
  {Deines-Jones}, {Esposito}, {Fichtel}, {Friedlander}, {Hunter}, {McDonald},
  {Sreekumar}, {Thompson}, {Jones}, {Lin}, {Michelson}, {Nolan}, {Tompkins},
  {Kanbach}, {Mayer-Hasselwander}, {M{\"u}cke}, {Pohl}, {Reimer}, {Kniffen},
  {Schneid}, {von Montigny}, {Mukherjee}, \& {Dingus}}]{Hartman1999}
{Hartman}, R.~C., {Bertsch}, D.~L., {Bloom}, S.~D., {et~al.} 1999, \apjs, 123,
  79

\bibitem[{Heidt \& Wagner(1996)}]{Heidt1996}
Heidt, J., \& Wagner, S.~J. 1996, \aap, 305, 42

\bibitem[{Homayouni {et~al.}(2019)Homayouni, Trump, Grier, Shen, Starkey,
  Brandt, Fonseca~Alvarez, Hall, Horne, Kinemuchi, I-Hsiu~Li, McGreer, Sun, Ho,
  \& Schneider}]{Homayouni2019}
Homayouni, Y., Trump, J.~R., Grier, C.~J., {et~al.} 2019, \apj, 880, 126

\bibitem[{{Isler} {et~al.}(2017){Isler}, {Urry}, {Coppi}, {Bailyn}, {Brady},
  {MacPherson}, {Buxton}, \& {Hasan}}]{Isler2017}
{Isler}, J.~C., {Urry}, C.~M., {Coppi}, P., {et~al.} 2017, \apj, 844, 107

\bibitem[{Jayasinghe {et~al.}(2019)Jayasinghe, Stanek, Kochanek, Shappee,
  Holoien, Thompson, Prieto, Dong, Pawlak, Pejcha, Shields, Pojmanski, Otero,
  Hurst, Britt, \& Will}]{Jayasinghe2019}
Jayasinghe, T., Stanek, K.~Z., Kochanek, C.~S., {et~al.} 2019, \mnras, 485, 961

\bibitem[{Jorstad {et~al.}(2001)Jorstad, Marscher, Mattox, Wehrle, Bloom, \&
  Yurchenko}]{Jorstad2001}
Jorstad, S.~G., Marscher, A.~P., Mattox, J.~R., {et~al.} 2001, \apjs, 134, 181.
\newblock \url{https://ui.adsabs.harvard.edu/abs/2001ApJS..134..181J}

\bibitem[{Kirk {et~al.}(1998)Kirk, Rieger, \& Mastichiadis}]{Kirk1998}
Kirk, J.~G., Rieger, F.~M., \& Mastichiadis, A. 1998, \aap, 333, 452

\bibitem[{Li {et~al.}(2019)Li, Shen, Brandt, Grier, Hall, Ho, Homayouni, Horne,
  Schneider, Trump, \& Starkey}]{Li2019}
Li, I-Hsiu, J., Shen, Y., Brandt, W.~N., {et~al.} 2019, \apj, 884, 119

\bibitem[{{Liodakis} {et~al.}(2019){Liodakis}, {Romani}, {Filippenko},
  {Kocevski}, \& {Zheng}}]{2019ApJ...880...32L}
{Liodakis}, I., {Romani}, R.~W., {Filippenko}, A.~V., {Kocevski}, D., \&
  {Zheng}, W. 2019, \apj, 880, 32

\bibitem[{MacLeod {et~al.}(2010)MacLeod, Ivezi{\'c}, Kochanek, Koz{\l}owski,
  Kelly, Bullock, Kimball, Sesar, Westman, Brooks, Gibson, Becker, \&
  de~Vries}]{MacLeod2010}
MacLeod, C.~L., Ivezi{\'c}, {\v{Z}}., Kochanek, C.~S., {et~al.} 2010, \apj,
  721, 1014

\bibitem[{Marscher(2013)}]{Marscher2013}
Marscher, A.~P. 2013, in European Physical Journal Web of Conferences, Vol.~61,
  European Physical Journal Web of Conferences, 04001.
\newblock \url{https://ui.adsabs.harvard.edu/abs/2013EPJWC..6104001M}

\bibitem[{Marscher \& Gear(1985)}]{Marscher1985}
Marscher, A.~P., \& Gear, W.~K. 1985, \apj, 298, 114

\bibitem[{Meng {et~al.}(2017)Meng, Wu, Webb, Zhang, \& Dai}]{Meng2017}
Meng, N., Wu, J., Webb, J.~R., Zhang, X., \& Dai, Y. 2017, \mnras, 469, 3588

\bibitem[{{Meng} {et~al.}(2018){Meng}, {Zhang}, {Wu}, {Ma}, \&
  {Zhou}}]{Meng2018}
{Meng}, N., {Zhang}, X., {Wu}, J., {Ma}, J., \& {Zhou}, X. 2018, \apjs, 237, 30

\bibitem[{Nanci {et~al.}(2022)Nanci, Giroletti, Orienti, Migliori, Mold{\'o}n,
  Garrappa, Kadler, Ros, Buson, An, P{\'e}rez-Torres, D'Ammando, Mohan, Agudo,
  Sohn, Castro-Tirado, \& Zhang}]{Nanci2022}
Nanci, C., Giroletti, M., Orienti, M., {et~al.} 2022, arXiv e-prints,
  arXiv:2203.13268.
\newblock \url{https://ui.adsabs.harvard.edu/abs/2022arXiv220313268N}

\bibitem[{Nemmen {et~al.}(2012)Nemmen, Georganopoulos, Guiriec, Meyer, Gehrels,
  \& Sambruna}]{Nemmen2012}
Nemmen, R.~S., Georganopoulos, M., Guiriec, S., {et~al.} 2012, Science, 338,
  1445.
\newblock \url{https://ui.adsabs.harvard.edu/abs/2012Sci...338.1445N}

\bibitem[{{Nolan} {et~al.}(2003){Nolan}, {Tompkins}, {Grenier}, \&
  {Michelson}}]{Nolan2003}
{Nolan}, P.~L., {Tompkins}, W.~F., {Grenier}, I.~A., \& {Michelson}, P.~F.
  2003, \apj, 597, 615

\bibitem[{Paltani {et~al.}(1997)Paltani, Courvoisier, Blecha, \&
  Bratschi}]{Paltani1997}
Paltani, S., Courvoisier, T. J.~L., Blecha, A., \& Bratschi, P. 1997, \aap,
  327, 539.
\newblock \url{https://ui.adsabs.harvard.edu/abs/1997A&A...327..539P}

\bibitem[{Pandey {et~al.}(2020)Pandey, Gupta, Kurtanidze, Wiita, Damljanovic,
  Bachev, Zhang, Kurtanidze, Darriba, Chigladze, Latev, Nikolashvili, Peneva,
  Semkov, Strigachev, Tiwari, \& Vince}]{Pandey2020}
Pandey, A., Gupta, A.~C., Kurtanidze, S.~O., {et~al.} 2020, \apj, 890, 72

\bibitem[{Perlman {et~al.}(2008)Perlman, Addison, Georganopoulos, Wingert, \&
  Graff}]{Perlman2008}
Perlman, E., Addison, B., Georganopoulos, M., Wingert, B., \& Graff, P. 2008,
  in Blazar Variability across the Electromagnetic Spectrum, 9

\bibitem[{Peterson {et~al.}(1998{\natexlab{a}})Peterson, Wanders, Bertram,
  Hunley, Pogge, \& Wagner}]{Peterson1998}
Peterson, B.~M., Wanders, I., Bertram, R., {et~al.} 1998{\natexlab{a}}, \apj,
  501, 82

\bibitem[{Peterson {et~al.}(1998{\natexlab{b}})Peterson, Wanders, Horne,
  Collier, Alexander, Kaspi, \& Maoz}]{Peterson1998a}
Peterson, B.~M., Wanders, I., Horne, K., {et~al.} 1998{\natexlab{b}}, \pasp,
  110, 660

\bibitem[{Peterson {et~al.}(2004)Peterson, Ferrarese, Gilbert, Kaspi, Malkan,
  Maoz, Merritt, Netzer, Onken, Pogge, Vestergaard, \& Wandel}]{Peterson2004}
Peterson, B.~M., Ferrarese, L., Gilbert, K.~M., {et~al.} 2004, \apj, 613, 682

\bibitem[{Polednikova {et~al.}(2016)Polednikova, Ederoclite, de~Diego, Cepa,
  Gonz{\'a}lez-Serrano, Bongiovanni, Oteo, Garc{\'\i}a,
  P{\'e}rez-Mart{\'\i}nez, Pintos-Castro, Ram{\'o}n-P{\'e}rez, \&
  S{\'a}nchez-Portal}]{Polednikova2016}
Polednikova, J., Ederoclite, A., de~Diego, J.~A., {et~al.} 2016, \mnras, 460,
  3950

\bibitem[{Qian {et~al.}(1991)Qian, Quirrenbach, Witzel, Krichbaum, Hummel, \&
  Zensus}]{Qian1991}
Qian, S.~J., Quirrenbach, A., Witzel, A., {et~al.} 1991, \aap, 241, 15

\bibitem[{{Rajput} {et~al.}(2020){Rajput}, {Stalin}, \&
  {Sahayanathan}}]{2020MNRAS.498.5128R}
{Rajput}, B., {Stalin}, C.~S., \& {Sahayanathan}, S. 2020, \mnras, 498, 5128

\bibitem[{{Rani} {et~al.}(2010){Rani}, {Gupta}, {Strigachev}, {Bachev},
  {Wiita}, {Semkov}, {Ovcharov}, {Mihov}, {Boeva}, {Peneva}, {Spassov},
  {Tsvetkova}, {Stoyanov}, \& {Valcheva}}]{Rani2010}
{Rani}, B., {Gupta}, A.~C., {Strigachev}, A., {et~al.} 2010, \mnras, 404, 1992

\bibitem[{{Rector} \& {Stocke}(2001)}]{Rector2001}
{Rector}, T.~A., \& {Stocke}, J.~T. 2001, \aj, 122, 565

\bibitem[{Safna {et~al.}(2020)Safna, Stalin, Rakshit, \& Mathew}]{Safna2020}
Safna, P.~Z., Stalin, C.~S., Rakshit, S., \& Mathew, B. 2020, \mnras, 498, 3578

\bibitem[{Sagar {et~al.}(2004)Sagar, Stalin, Gopal-Krishna, \&
  Wiita}]{Sagar2004}
Sagar, R., Stalin, C.~S., Gopal-Krishna, \& Wiita, P.~J. 2004, \mnras, 348, 176

\bibitem[{Sahakyan {et~al.}(2022)Sahakyan, Giommi, Padovani, Petropoulou,
  B{\'e}gu{\'e}, Boccardi, \& Gasparyan}]{Sahakyan2022}
Sahakyan, N., Giommi, P., Padovani, P., {et~al.} 2022, arXiv e-prints,
  arXiv:2204.05060.
\newblock \url{https://ui.adsabs.harvard.edu/abs/2022arXiv220405060S}

\bibitem[{{Sandrinelli} {et~al.}(2014){Sandrinelli}, {Covino}, \&
  {Treves}}]{Sandrinelli2014}
{Sandrinelli}, A., {Covino}, S., \& {Treves}, A. 2014, \aap, 562, A79

\bibitem[{Scarpa {et~al.}(2000)Scarpa, Urry, Padovani, Calzetti, \&
  O'Dowd}]{Scarpa2000}
Scarpa, R., Urry, C.~M., Padovani, P., Calzetti, D., \& O'Dowd, M. 2000, \apj,
  544, 258

\bibitem[{Shah {et~al.}(2021)Shah, Jithesh, Sahayanathan, \& Iqbal}]{Shah2021}
Shah, Z., Jithesh, V., Sahayanathan, S., \& Iqbal, N. 2021, \mnras, 2103.13657

\bibitem[{Shah {et~al.}(2019)Shah, Jithesh, Sahayanathan, Misra, \&
  Iqbal}]{Shah2019}
Shah, Z., Jithesh, V., Sahayanathan, S., Misra, R., \& Iqbal, N. 2019, \mnras,
  484, 3168

\bibitem[{{Sikora} {et~al.}(2009){Sikora}, {Stawarz}, {Moderski}, {Nalewajko},
  \& {Madejski}}]{Sikora2009}
{Sikora}, M., {Stawarz}, {\L}., {Moderski}, R., {Nalewajko}, K., \& {Madejski},
  G.~M. 2009, \apj, 704, 38

\bibitem[{Simonetti {et~al.}(1985)Simonetti, Cordes, \&
  Heeschen}]{Simonetti1985}
Simonetti, J.~H., Cordes, J.~M., \& Heeschen, D.~S. 1985, \apj, 296, 46.
\newblock \url{https://ui.adsabs.harvard.edu/abs/1985ApJ...296...46S}

\bibitem[{Smith {et~al.}(2009)Smith, Montiel, Rightley, Turner, Schmidt, \&
  Jannuzi}]{Smith2009}
Smith, P.~S., Montiel, E., Rightley, S., {et~al.} 2009, arXiv e-prints,
  arXiv:0912.3621

\bibitem[{Sun {et~al.}(2018)Sun, Grier, \& Peterson}]{Sun2018}
Sun, M., Grier, C.~J., \& Peterson, B.~M. 2018, PyCCF: Python Cross Correlation
  Function for reverberation mapping studies, , , ascl:1805.032

\bibitem[{Tody(1986)}]{Tody1986}
Tody, D. 1986, in Society of Photo-Optical Instrumentation Engineers (SPIE)
  Conference Series, Vol. 627, Instrumentation in astronomy VI, ed. D.~L.
  {Crawford}, 733

\bibitem[{Tody(1993)}]{Tody1993}
Tody, D. 1993, in Astronomical Society of the Pacific Conference Series,
  Vol.~52, Astronomical Data Analysis Software and Systems II, ed. R.~J.
  {Hanisch}, R.~J.~V. {Brissenden}, \& J.~{Barnes}, 173

\bibitem[{{Ulrich} {et~al.}(1997){Ulrich}, {Maraschi}, \& {Urry}}]{Ulrich1997}
{Ulrich}, M.-H., {Maraschi}, L., \& {Urry}, C.~M. 1997, \araa, 35, 445

\bibitem[{{Urry} \& {Padovani}(1995)}]{Urry1995}
{Urry}, C.~M., \& {Padovani}, P. 1995, \pasp, 107, 803

\bibitem[{Virtanen \& Vainio(2005)}]{Virtanen2005}
Virtanen, J. J.~P., \& Vainio, R. 2005, \apj, 621, 313

\bibitem[{Wang \& Jiang(2020)}]{Wang2020}
Wang, Y.-F., \& Jiang, Y.-G. 2020, \apj, 902, 41

\bibitem[{Weaver {et~al.}(2022)Weaver, Jorstad, Marscher, Morozova, Troitsky,
  Agudo, G{\'o}mez, L{\"a}hteenm{\"a}ki, Tammi, \& Tornikoski}]{Weaver2022}
Weaver, Z.~R., Jorstad, S.~G., Marscher, A.~P., {et~al.} 2022, \apjs, 260, 12.
\newblock \url{https://ui.adsabs.harvard.edu/abs/2022ApJS..260...12W}

\bibitem[{{Webb} {et~al.}(1988){Webb}, {Smith}, {Leacock}, {Fitzgibbons},
  {Gombola}, \& {Shepherd}}]{Webb1988}
{Webb}, J.~R., {Smith}, A.~G., {Leacock}, R.~J., {et~al.} 1988, \aj, 95, 374

\bibitem[{Webb {et~al.}(1998)Webb, Freedman, Howard, Ma, Belfort, Rave,
  Rumstay, Nicol, Krick, Oswalt, Marshall, \& Robishaw}]{Webb1998}
Webb, J.~R., Freedman, I., Howard, E., {et~al.} 1998, \aj, 115, 2244

\bibitem[{Weng {et~al.}(2020)Weng, Chen, Wang, Cai, Qiao, \& Liao}]{Weng2020}
Weng, S.-S., Chen, Y., Wang, T.-T., {et~al.} 2020, \mnras, 491, 2576

\bibitem[{{Williamson} {et~al.}(2014){Williamson}, {Jorstad}, {Marscher},
  {Larionov}, {Smith}, {Agudo}, {Arkharov}, {Blinov}, {Casadio}, {Efimova},
  {G{\'o}mez}, {Hagen-Thorn}, {Joshi}, {Konstantinova}, {Kopatskaya},
  {Larionova}, {Larionova}, {Malmrose}, {McHardy}, {Molina}, {Morozova},
  {Schmidt}, {Taylor}, \& {Troitsky}}]{Williamson2014}
{Williamson}, K.~E., {Jorstad}, S.~G., {Marscher}, A.~P., {et~al.} 2014, \apj,
  789, 135

\bibitem[{{Wu} {et~al.}(2012){Wu}, {B{\"o}ttcher}, {Zhou}, {He}, {Ma}, \&
  {Jiang}}]{Wu2012}
{Wu}, J., {B{\"o}ttcher}, M., {Zhou}, X., {et~al.} 2012, \aj, 143, 108

\bibitem[{{Wu} {et~al.}(2011){Wu}, {Zhou}, {Ma}, \& {Jiang}}]{Wu2011}
{Wu}, J., {Zhou}, X., {Ma}, J., \& {Jiang}, Z. 2011, \mnras, 418, 1640

\bibitem[{Wu {et~al.}(2007)Wu, Zhou, Ma, Wu, Jiang, \& Chen}]{Wu2007}
Wu, J., Zhou, X., Ma, J., {et~al.} 2007, \aj, 133, 1599

\bibitem[{Xue \& Cui(2005)}]{Xue2005}
Xue, Y., \& Cui, W. 2005, \apj, 622, 160

\bibitem[{{Yan} {et~al.}(2013){Yan}, {Zhang}, {Yuan}, {Fan}, \&
  {Zeng}}]{2013ApJ...765..122Y}
{Yan}, D., {Zhang}, L., {Yuan}, Q., {Fan}, Z., \& {Zeng}, H. 2013, \apj, 765,
  122

\bibitem[{{Yuan} \& {Fan}(2021)}]{2021PASP..133g4101Y}
{Yuan}, Y.~H., \& {Fan}, J.~H. 2021, \pasp, 133, 074101

\bibitem[{Zhang {et~al.}(2018)Zhang, Wu, \& Meng}]{Zhang2018}
Zhang, X., Wu, J., \& Meng, N. 2018, \mnras, 478, 3513

\bibitem[{Zu {et~al.}(2013)Zu, Kochanek, Koz{\l}owski, \& Udalski}]{Zu2013}
Zu, Y., Kochanek, C.~S., Koz{\l}owski, S., \& Udalski, A. 2013, \apj, 765, 106

\bibitem[{Zu {et~al.}(2011)Zu, Kochanek, \& Peterson}]{Zu2011}
Zu, Y., Kochanek, C.~S., \& Peterson, B.~M. 2011, \apj, 735, 80

\end{thebibliography}

\end{document}